\documentclass[12pt,onecolumn,twoside]{IEEEtran}
\usepackage[utf8]{inputenc}
\usepackage[T1]{fontenc}
\usepackage{url}              % provides \url{...}
\usepackage{cite}             % improves presentation of citations

\usepackage[cmex10]{amsmath}  % Use the [cmex10] option to ensure complicance
\usepackage{mleftright}       % fix to wrong spacing of \left-,
\mleftright                   % \middle- \right-commands

\usepackage{graphicx}         % provides \includegraphics{...} to
\usepackage{booktabs}         % fixes poor spacing in tables and
\usepackage{amsmath}
\usepackage{latexsym}
\usepackage{amssymb}

\newcommand{\cA}{{\cal A}}
\newcommand{\cB}{{\cal B}}
\newcommand{\cC}{{\cal C}}
\newcommand{\cD}{{\cal D}}

\newcommand{\cF}{{\cal F}}

\newcommand{\cS}{{\cal S}}
\newcommand{\cT}{{\cal T}}

\newcommand{\cQ}{{\cal Q}}

\newcommand{\cX}{{\cal X}}
\newcommand{\cY}{{\cal Y}}

\newcommand{\bD}{{\bf D}}
\newcommand{\bE}{{\bf E}}

\newcommand{\bfone}{{\bf 1}}

\newcommand{\mmod}{{\mbox{mod}}}

\newcommand{\field}[1]{\mathbb{#1}}

\newcommand{\C}{\field{C}}

\newcommand{\F}{\field{F}}

\newcommand{\Z}{\field{Z}}

\newtheorem{theorem}{Theorem}
\newtheorem{definition}{Definition}
\newtheorem{lemma}{Lemma}
\newtheorem{corollary}{Corollary}
\newtheorem{example}{Example}
\newtheorem{construction}{Construction}
\newtheorem{conjecture}{Conjecture}

\begin{document}

\title{On de Bruijn Array Codes,\\Part I: Nonlinear Codes}

%%%%%%
\author{%
  \IEEEauthorblockN{Tuvi Etzion}
%  \IEEEauthorblockA{%
%    Please do NOT provide authors' names and affiliations\\
%    in the paper submitted for review, but keep this placeholder.\\
%    ISIT23 follows a \textbf{double-blind reviewing policy}.}
}

\maketitle

%%%%%
%% Abstract:
%% If your paper is eligible for the student paper award, please add
%% the comment "THIS PAPER IS ELIGIBLE FOR THE STUDENT PAPER
%% AWARD." as a first line in the abstract.
%% For the final version of the accepted paper, please do not forget
%% to remove this comment!
%%
\begin{abstract}
A de Bruijn array code is a set of $r \times s$ binary doubly-periodic arrays such that each binary
$n \times m$ matrix is contained exactly once as a window in one of the arrays.
Such a set of arrays can be viewed as a two-dimensional generalization of a perfect factor
in the de Bruijn graph. Necessary conditions for the existence of such codes are given.
Several direct constructions and recursive constructions for such arrays are given. A framework for a theory
of two-dimensional feedback shift registers which is akin to (one-dimensional) feedback shift registers is suggested in the process.
\end{abstract}

\section{Introduction}
\label{sec:PM+PR}

Generalizations of one-dimensional sequences and codes to higher dimensions are quite natural from
both theoretical and practical points of view. Such generalizations were considered
for various structures such as error-correcting codes~\cite{BlBr00,Rot91}, burst-correcting codes~\cite{BBV,BBZS,EtVa02,EtYa09},
constrained codes~\cite{ScBr08,TER09}, and de Bruijn sequences~\cite{Etz88,Mit95,Pat94}.
This paper considers a generalization of one-dimensional sequences with a window property to two-dimensional arrays with a window
property. For simplicity in this paper only binary arrays and usually binary sequences are considered, although some of the
results can be generalized to any alphabet size and some sequences over larger alphabets can be used to construct
binary arrays.

The de Bruijn graph $G_{\sigma,n} =(V_n,E_n)$ ($G_n$ when $\sigma=2$) is a directed graph, where the set of vertices~$V_n$
has $\sigma^n$ vertices represented by the set of $\sigma^n$ words of length $n$ over an alphabet of size $\sigma$.
The set of edges $E_n$ consists of $\sigma^{n+1}$ directed edges represented by the $\sigma^{n+1}$ words of length $n+1$.
The edge $(x_0,x_1,\ldots,x_{n-1},x_n)$ is directed from the vertex $(x_0,x_1,\ldots,x_{n-1})$ to the vertex $(x_1,\ldots,x_{n-1},x_n)$.
A~{\bf \emph{span~$n$ de Bruijn sequence}} is a cyclic sequence of length $\sigma^n$ in which each $n$-tuple is contained in
exactly one window of $n$ consecutive digits. This cyclic sequence is equivalent to an Eulerian cycle in $G_{\sigma,n-1}$ and
a Hamiltonian cycle in $G_{\sigma,n}$. A {\bf \emph{span $n$ shortened de Bruijn sequence}} is a sequence of length $\sigma^n-1$ in which
each nonzero $n$-tuple is contained in exactly one window of $n$ consecutive digits.
The following definition generalizes the definition of a de Bruijn sequence to a two-dimensional array.

\begin{definition}
A {\bf \emph{de Bruijn array}} (often known as a {\bf \emph{perfect map}}) is an $r \times s$ doubly-periodic array
(cyclic horizontally and vertically like in a torus),
such that each $n \times m$ binary matrix appears exactly once as a window in the array.
\end{definition}

Perfect maps were first presented by Reed and Stewart~\cite{ReSt62} and later in Gordon~\cite{Gor66} and Clapham~\cite{Cla86}. A~very simple
construction based on a span $n$ de Bruijn sequence and a de Bruijn sequence over an alphabet of size $2^n$ was presented
by Ma~\cite{Ma84}. This construction will be modified later in one of our constructions.
Fan, Fan, Ma, and Siu~\cite{FFMS85} gave a two-dimensional graphical representation for perfect maps.
They also presented two recursive constructions for such de Bruijn arrays. An important construction for perfect maps and
another graphical representation was given in~\cite{Etz88}. Finally, Paterson~\cite{Pat94} proved that the
necessary conditions for the existence of de Bruijn arrays are also sufficient.

These arrays have found a variety of applications.
They were used in pattern recognition for structured light systems as described in Geng~\cite{Gen11},
Lin, Nie, and Song~\cite{LNS16}, Morano, Ozturk, Conn, Dubin, Zietz, and Nissanov~\cite{MOCDZN98},
Salvi, Fernandez, Pribanic, and Llado~\cite{SFPL10}, and Salvi, Pag\`{e}s, and Batlle~\cite{SPB04}.
They are also used in transferring planar surface into a sensitive touch screen display, see Dai and Chung~\cite{DaCh14}, in
camera localization as described by Szentandrasi, Zachari\'{a}\u{s}, Havel, Herout, Dubska, and Kajan~\cite{SZHHDK}, and
in one-shot shape acquisition, e.g., Pag\`{e}s, Salvi, Collewet, and Forest~\cite{PSCF05}. Finally, they can be applied
to surface measurements as described in Kiyasu, Hoshino, Yano, and Fujimura~\cite{KHYF95} and in Spoelder, Vos, Petriu, and Groen~\cite{SVPG00},
and also in coded aperture imaging as was presented for example in Gottesman and Fenimore~\cite{GoFe89}.

The de Bruijn arrays have been extensively covered in the literature.
Further work on this topic and related structures was done by~\cite{HuIs93,HuIs95,HMP96,LMS79,McSl76,Mit94,Mit95,Mit97,MiPa94,MiPa98}

The goal of the current paper is to extend the results
on de Bruijn arrays into de Bruijn array codes defined as follows.

\begin{definition}
A {\bf \emph{de Bruijn array code}} is a set of $r \times s$ doubly-periodic arrays,
such that each nonzero $n \times m$ matrix appears exactly once as a window in one of the arrays.
Such a set of arrays will be referred to as $(r,s;n,m)$-DBAC.
The size $\Delta$ of an array code is the number of arrays (codewords) in the code.
\end{definition}
It is straightforward to verify the necessary conditions for the existence of an $(r,s;n,m)$-DBAC.
\begin{lemma}
\label{lem:condPMC}
If $\C$ is an $(r,s;n,m)$-$\textup{DBAC}$ of size $\Delta$, then
\begin{enumerate}
\item[{\bf 1.}] $r > n$ or $r=n=1$,

\item[{\bf 2.}] $s > m$ or $s=m=1$.

\item[{\bf 3.}] $\Delta rs = 2^{nm}$.
\end{enumerate}
\end{lemma}
Lemma~\ref{lem:condPMC} implies that $r$, $s$, and $\Delta$, are powers of two. Henceforth, we use
$2^k$ instead of $r$ and $2^t$ instead of $s$.
One of the goals in the research on DBACs is to prove that the necessary conditions of Lemma~\ref{lem:condPMC} are sufficient.
In the discussion given here it will also be shown how the definition
of de Bruijn array codes generalizes some one-dimensional definitions.
For a given $(2^k,2^t;n,m)$-BBAC we can always assume that $2 \leq k,n$ and $1 \leq m,t$. Further restriction associated
with some constructions will be given later.

The rest of this paper is organized as follows. In Section~\ref{sec:preliminaries} we present the basic definitions and theory for
the arrays considered here. In particular the section introduces some theory of shift-register sequences and factors (state diagrams)
of the de Bruijn graph.
Many constructions are presented in the paper, some of these constructions are
direct constructions and some of them are recursive constructions. Section~\ref{sec:direct_construct} is devoted
to a few direct constructions.
Recursive constructions will be presented in Section~\ref{sec:recursive_construct}.
Perfect factors have to be used carefully for these constructions and the chosen arrays on which the
constructions are applied should be also chosen in an appropriate way.
In Section~\ref{sec:join} another technique to generate $(2^k,2^t;n,m)$-DBACs by joining codewords (cycles) of a $(2^k,2^{t'};n,m)$-DBAC with
shorter codewords, i.e., $t' < t$, is presented. This technique can be applied if all the codewords are cycles in the same
state diagram of the two-dimensional generalization de Bruijn graph.
Section~\ref{sec:analysis} contains an analysis of the constructions and their input. It also summarizes of some parameters of
$(2^k,2^t;n,m)$-DBACs obtained here. Section~\ref{sec:conclusion} includes conclusions and future work.

\section{Preliminaries}
\label{sec:preliminaries}

\subsection{Shift-register sequences}
\label{sec:FSR}

\begin{figure}[ht]
\vspace{0.4cm}
\begin{picture}(115,115)(-70,50)
\includegraphics[width=10cm]{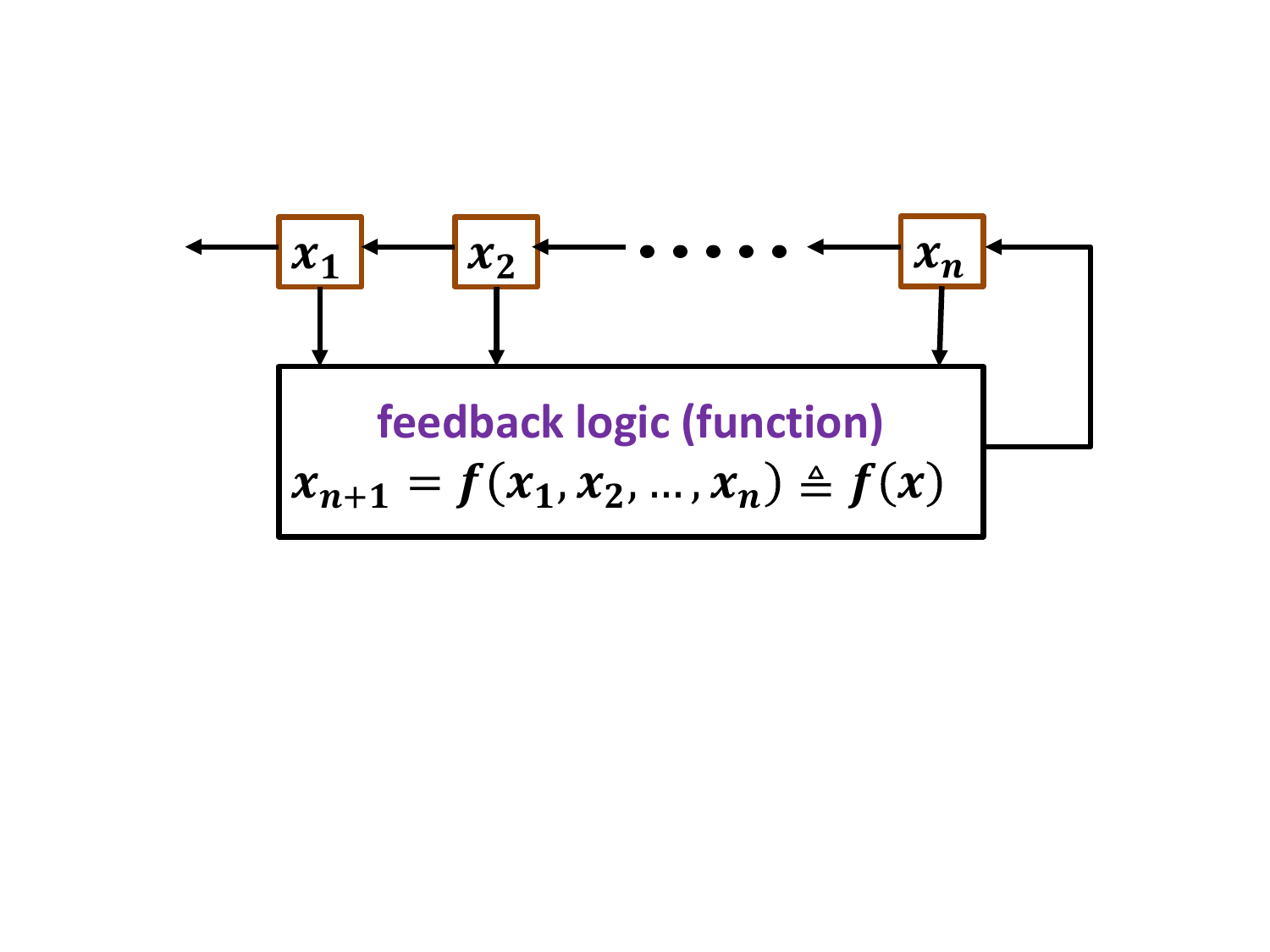}
\end{picture}
\vspace{-0.3cm}
\caption{Feedback shift register of order $n$.}
\label{fig:FSRn}
\end{figure}

A {\bf \emph{feedback shift register}} of order $n$ (an FSR$_n$ in short) has $2^n$ {\bf \emph{states}}, represented by the set
of $2^n$~binary words of length $n$.
A register has $n$ cells, which are binary storage elements,
where each cell stores at each stage one of the bits of the current state $x=(x_1,x_2,\ldots,x_n)$.
An FSR$_n$ is depicted in Fig.~\ref{fig:FSRn}.

Shift registers and their sequences have been extensively studied and three comprehensive books on them are~\cite{Etz24,Gol67,GoKl12}.
The material in this section is presented in these books.
If the word $(x_1,x_2, \ldots , x_n)$ is the state of the FSR$_n$, then $x_i$ is stored in the $i$-th cell of the FSR$_n$.
The $n$ cells are connected to another logic element which computes a Boolean feedback function
$f(x_1,x_2,\ldots,x_n)$. At periodic intervals, controlled by a master clock, $x_2$ is transferred to $x_1$,
$x_3$ to $x_2$, and so on until $x_n$ is transferred to $x_{n-1}$. The value of the feedback function is
transferred to $x_n$ and hence it is common to write $x_{n+1}=f(x_1,x_2,\ldots,x_n)$.
The register starts with an {\bf \emph{initial state}} $(a_1,a_2,\ldots,a_n)$, where $a_i$, $1 \leq i \leq n$,
is the initial value stored in the $i$-th cell.
There are $2^n$ possible distinct values for $x_1,x_2,\ldots,x_n$, i.e., there are $2^n$~distinct states and each one can have
a value of either 0 or 1. Hence, there are $2^{2^n}$ different FSR$_n$, but not all of these functions are of interest here.

Each FSR$_n$ has a {\bf \emph{state diagram}}, a graph with $2^n$ vertices (the states of the FSR$_n$).
Given an FSR$_n$ with feedback function
$f(x_1,x_2,\ldots,x_n)$, the vertex $(x_1,x_2,\ldots,x_n)$ in the associated state diagram has an edge to the
vertex $(x_2,\ldots,x_n,x_{n+1})$ if $x_{n+1}=f(x_1,x_2,\ldots,x_n)$. This implies that the state diagram with
$2^n$ states (vertices) has exactly $2^n$ edges.
A {\bf \emph{nonsingular feedback shift register}} has a feedback function whose
state diagram contains only cycles.
A {\bf \emph{factor}} in a graph is a set of vertex-disjoint cycles that contain all the vertices in the graph;
a factor in $G_n$ is associated with a state diagram of a nonsingular feedback shift register.
These are the only FSR$_n$s which are of interest here.
%and hence in the rest of the paper all FSR$_n$s are nonsingular.
%We can easily verify if a feedback shift register is nonsingular using the following lemma.
%
%\begin{lemma}
%\label{lem:nonsingular_neq}
%An $\textup{FSR}_n$ is nonsingular if and only if for each $x_i \in \{ 0,1 \}$, $2 \leq i \leq n$,
%we have
%$$
%f(0,x_2,\ldots,x_n) \neq f(1,x_2,\ldots,x_n)
%$$
%\end{lemma}
%
%\begin{lemma}
%\label{lem:nonsingular_neq}
%An $\textup{FSR}_n$ is nonsingular if and only if for each $x_i \in \{ 0,1 \}$, $2 \leq i \leq n$,
%we have
%$$
%f(x_1,x_2,\ldots,x_n) = x_1 + g(x_2,\ldots,x_n) ~,
%$$
%where $g(x_2,\ldots,x_n)$ is any Boolean function with $n-1$ variables.
%\end{lemma}
%
%A nonsingular {\bf \emph{linear feedback shift register}} of order $n$ (an LFSR$_n$ in short) over $\F_2$ is an FSR$_n$
%whose feedback function $f$ is linear, i.e.,
%$$
%x_{n+1} = f(x_1,x_2,\ldots,x_n) = x_1 + \sum_{i=2}^n c_i x_i, ~~ c_i \in \F_2, ~~ 1 \leq i \leq n ~.
%$$
%
%The class of linear shift registers is very important, but most of the shift registers are nonlinear.
%While, by Lemma~\ref{lem:nonsingular_neq}, there are $2^{2^{n-1}}$ nonsingular feedback shift registers of order $n$,
%there are only $2^{n-1}$ nonsingular LFSR$_n$s.
%Some shift register feedback functions have properties similar
%to the linear functions, making them of particular interest here;
%two such classes of feedback function are of interest here.
Two classes of nonlinear FSRs are of interest here.
%
%The first class denoted by PCR$_n$ for the {\bf \emph{pure cycling register}}
%of order $n$, has linear feedback function
%$$
%f(x_1,x_2,\ldots,x_n)=x_1 ~.
%$$

The first class denoted by CCR$_n$ for the {\bf \emph{complemented cycling register}} of order $n$,
has the feedback function
$$
f(x_1,x_2,\ldots,x_n)=x_1 +1 ~.
$$

%The third class denoted by PSR$_n$ for the {\bf \emph{pure summing register}} of order~$n$,
%has linear feedback function
%$$
%f(x_1,x_2,\ldots,x_n)= \sum_{i=1}^n x_i ~.
%$$

The second class denoted by CSR$_n$ for the {\bf \emph{complemented summing register}} of order $n$,
has the feedback function
$$
f(x_1,x_2,\ldots,x_n)= \sum_{i=1}^n x_i +1~.
$$

The length of the cycles in the state diagrams of these shift-registers are of special interest.
The length of a cycle in the state diagram is the {\bf \emph{period}} of the associated sequence.
A {\bf cyclic} sequence will be denoted by $[s_0 s_1 \cdots s_{n-1}]$ and an {\bf acyclic} sequence (a word) by $(s_0 s_1 \cdots s_{n-1})$.
A {\bf \emph{doubly-periodic}} array is cyclic both horizontally and vertically.
The {\bf \emph{length}} of the sequence is the number of symbols in the sequence.
Hence, the length of the sequence $S=[s_0 s_1 \cdots s_{n-1}]$ is~$n$.
{\bf \emph{The period}} $\pi (S)$ of the cyclic sequence $\cS=[s_0 s_1 \cdots s_{n-1}]$,
is the least positive integer $\pi$ such that $s_i = s_{\pi +i}$,
for each $0 \leq i \leq n-1$, where indices are taken modulo $n$, when the sequence is cyclic.
%For a cyclic sequence $\cS=[s_0 s_1 \cdots s_{n-1}]$ any integer $\pi$, $1 \leq \pi \leq n$, such that
%$\pi$ divides $n$ and $s_i = s_{\pi+i}$ for $0 \leq i \leq n-1$ is {\bf \emph{a period}} of~$S$, but the period is the smallest such integer.
Two cyclic sequences are said to be {\bf \emph{equivalent}} if one is a cyclic shift of the other. Such sequences are usually considered
to be the same unless the starting point of the cyclic sequence is determined.
Similar definitions for the length and the period are applied to arrays.
A cycle of the state diagram can be represented by a sequence with length equal to the length of the cycle.
The sequence is generated from the first bits of the vertices in the cycle keeping their order as the order of the
vertices in the cycle. The period of a $2^k \times 2^t$ doubly-periodic array can be less than $2^k$ vertically and less than $2^t$
horizontally since doubly-periodic indicates only that the array is cyclic in both directions. However, this scenario
will not happened in the arrays which will be discussed in the paper.

%\begin{example}
%The cyclic sequence
%$$
%[010010010]
%$$
%is of length 9, but its period is 3. It also has periods 6 and 9.
%
%The cyclic sequence
%$$
%[00101]
%$$
%is of length 5 and its period is 5.
%
%The two sequences
%$$
%[0011101111001] ~~~ \text{and} ~~~ [1011110010011]
%$$
%are equivalent sequences as one is a cyclic shift of the other, i.e.,
%$$
%[0011101111001] \simeq [1011110010011].
%$$
%
%\hspace {17.2cm} $\blacksquare$
%\end{example}

The {\bf \emph{companion}} of a vertex (state) $x=(x_1,x_2,\ldots,x_{n-1},x_n)$
is the vertex $x'=(x_1,x_2,\ldots,x_{n-1},\bar{x}_n)$, where $\bar{b}$ is the binary complement of $x$.
The {\bf \emph{complement}} of a state $x=(x_1,x_2,\ldots,x_{n-1},x_n)$ is the
state $\bar{x}=(\bar{x}_1,\bar{x}_2,\ldots,\bar{x}_{n-1},\bar{x}_n)$. A cycle (sequence)
$\cS = [s_1,s_2,\ldots,s_k]$ is called {\bf \emph{self-dual}} if the complement of  $\cS$,
${\bar{\cS} = [\bar{s}_1,\bar{s}_2,\ldots,\bar{s}_k]}$ is equivalent to $\cS$.
A sequence $\cS$ is self-dual if and only if it has the form $\cS = [ X ~ \bar{X} ]$ , where the length of $\cS$ is also its period.

\begin{figure}[ht]
\vspace{0.4cm}
\begin{picture}(125,125)(-70,50)
\includegraphics[width=10cm]{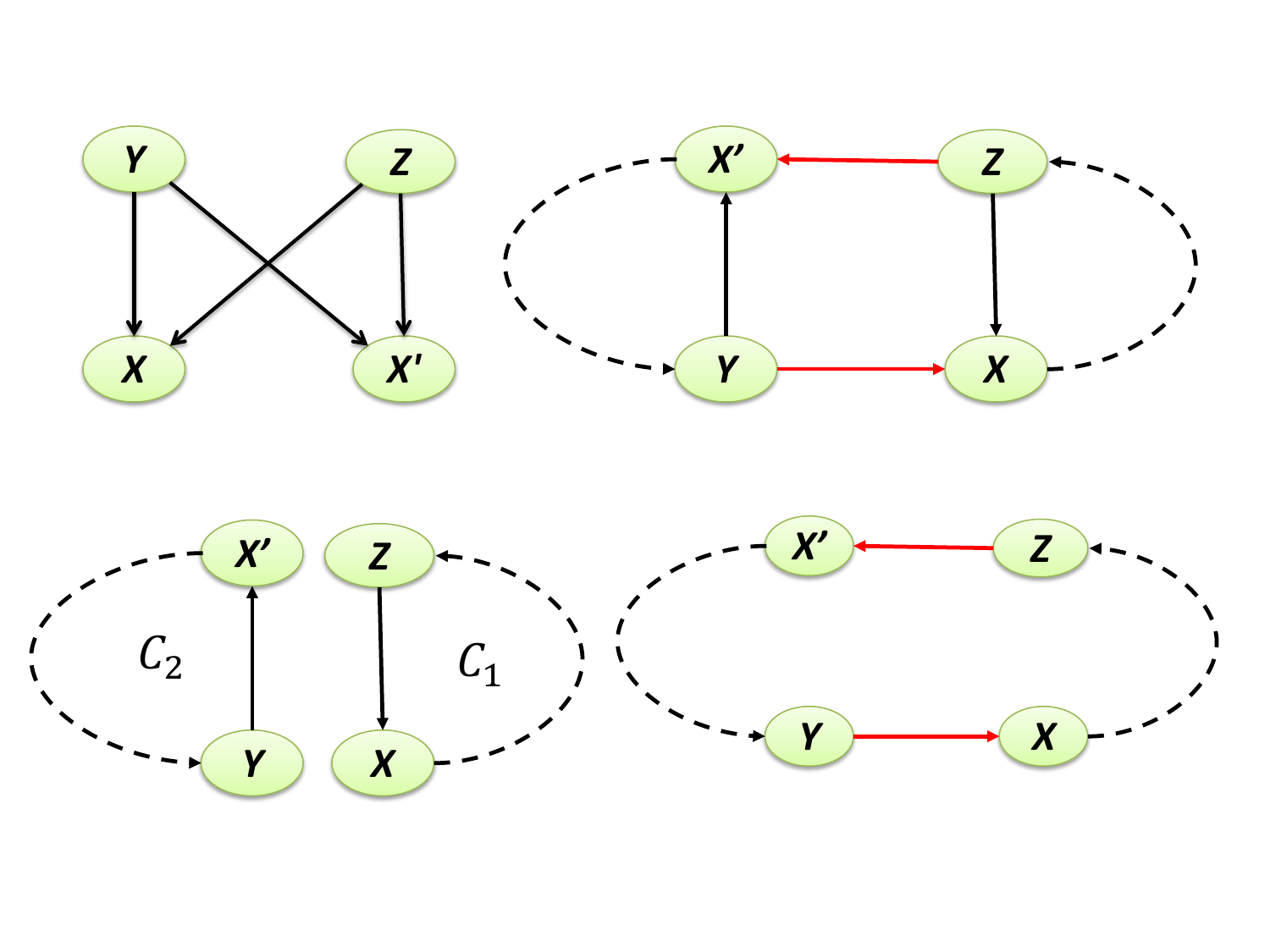}
\end{picture}
\vspace{0.45cm}
\caption{The merge-or-split method.}
\label{fig:cross-join}
\end{figure}

Two cycles $\cC_1$ and $\cC_2$ with a state $x$ in $\cC_1$ and its companion $x'$ in $\cC_2$ are merged into
one cycle $\cC$ when the predecessors of $x$ and $x'$, $z$ on $\cC_1$ and $y$ on $\cC_2$, respectively, are interchanged.
This operation is called the {\bf \emph{merge-or-split}} method since it either merges two cycles into one cycle or splits
one cycle into two cycles. The two possible scenarios of the merge-or-split method are depicted in Fig.~\ref{fig:cross-join}.

\subsection{The shift operator $\bE$}
\label{sec:Eoperator}

An operator with which many of the properties of feedback shift registers can
be represented is the shift operator $\bE$. This operator shifts the whole sequence to the left.
For a sequence $\cS= s_0 s_1 ~ \cdots ~ s_{k-1}$ of length $k$,
$\bE s_i = s_{i+1}$, where $0 \leq i < k-1$ and if the sequence is cyclic then also $\bE s_{k-1} = s_0$.
When the operator is applied to the cyclic sequence $\cS$ we have
$$
\bE [s_0 s_1 ~ \cdots ~ s_{k-1}] = [s_1 ~ \cdots ~ s_{k-1} s_0] ~.
$$
When the operator is applied on an acyclic sequence $\cS$ we have that
$$
\bE (s_0 s_1 ~ \cdots ~ s_{k-1}) = (s_1 ~ \cdots ~ s_{k-1}) ~.
$$
Any periodic (cyclic) sequence satisfies a linear recursion
\begin{equation}
\label{eq:lin_rec}
s_{i+m} = a_1 s_{i+m-1} + \cdots + a_{m-1} s_{i+1} + a_m s_i ~, ~  i \geq 0,
\end{equation}
of order $m \leq k$, where $a_j \in \{0,1\}$ and all computations are done modulo 2.
In terms of the shift operator~$\bE$ the linear recursion~(\ref{eq:lin_rec}) takes the form

\begin{equation}
\label{eq:lin_com}
\bE^m s_i = \sum_{j=1}^m a_j \bE^{m-j} s_i ~~ \text{or} ~~ \left(  \bE^m + \sum_{j=1}^m a_j \bE^{m-j} \right) s_i = 0~, ~~~~  i \geq 0.
\end{equation}
If $m$ is the smallest such integer, it implies that ${a_m \neq 0}$.
If we define
$$
f(\bE) \triangleq   \bE^m + \sum_{j=1}^m a_j \bE^{m-j} ~,
$$
then we have $f(\bE)s_i =0$ for each $i \geq 0$. This implies that~(\ref{eq:lin_com}) takes the form
$$
f(\bE)s_i = \left(  \bE^m + \sum_{j=0}^{m-1} c_j \bE^j \right) s_i = 0~,
$$
for each $i \geq 0$. In this case it is said that the polynomial $f(\bE)$ {\bf \emph{generates the sequence}} $\cS$
(or $\cS$ is generated by $f(\bE)$).

The {\bf \emph{linear complexity}} $c (\cS)$ of $\cS$ is defined as the least~$m$ for which~(\ref{eq:lin_rec}) holds.
Clearly ${c(\cS) \leq k}$ for any sequence with period $k$, since $s_{i+k}=s_i$ by definition.

The sequences that are of interest in our exposition have length a power of 2.
These sequences are generated by the polynomial $(\bE+1)^k$ and for each sequence $\cS$ generated by this
polynomial we have $(\bE+\bfone)^k \cS =[0,0,\ldots,0]$.
The following properties characterize these sequences~\cite{Etz24}.

\begin{lemma}
A sequence $\cS$ of length $2^n$ has linear complexity $c$ if and only if $(\bE +\bfone)^{c-1} \cS = [1^{2^n}]$.
\end{lemma}

\begin{lemma}
\label{lem:setsFor PF}
There are $2^{n - \lfloor \log n \rfloor -1}$ sequences which are generated by the polynomial $(\bE+ \bfone)^{n+1}$ and not generated by the
polynomial $(\bE+\bfone)^n$. All these sequences have period $2^{\lfloor \log n \rfloor +1}$ and linear complexity~${n+1}$.
Each $n$-tuple is contained exactly once in one of these sequences.
\end{lemma}

We define the {\bf \emph{weight}} of a word $(x_1,x_2,\ldots,x_n)$ as the number of nonzero entries in the word.
Similarly, we define the {\bf \emph{weight}} of a cyclic sequence $[x_1,x_2,\ldots,x_n]$ as the number of nonzero entries in the sequence.
Note, that the weight of the word $(011)$ is two, of the sequence $[011]$ is also two, but of the sequence $[011011]$ is four,
although its period is three and the weight of one period is two.
Other properties of sequences are as follows.
\begin{lemma}
For $n \geq 1$, a sequence $\cS$ of length $2^n$ has linear complexity $c >0$ if and only if the linear complexity
of the sequence $(\bE +\bfone) \cS$ is $c-1$.
\end{lemma}
\begin{lemma}
\label{lem:oddW_comp}
For $n \geq 1$, a sequence $\cS$ of length $2^n$ has linear complexity $2^n$ if and only if its weight is odd.
\end{lemma}
\begin{lemma}
\label{lem:comp_SD}
For $n \geq 1$, a sequence $\cS$ of length $2^{n+1}$ has linear complexity $2^n+1$ if and only if $\cS$ is a self-dual sequence.
\end{lemma}

Finally, we consider the linear complexity of de Bruijn sequences~\cite{CGK82,Etz99,EtLe84a,EtLe84b,Gam83}.
\begin{theorem}
\label{thm:deBcomp}
For $n \geq 3$, if $\cS$ is a span $n$ de Bruijn sequence, then $2^{n-1} +n \leq c(\cS) \leq 2^n-1$ and $c(\cS) \neq 2^{n-1}+n+1$.
There exists a span $n$ de Bruijn sequence $\cS$ for which $c(\cS)=2^{n-1}+n$ and a span~$n$ de Bruijn sequence $\cS$ for which $c(\cS)=2^n-1$.
\end{theorem}

\begin{conjecture}
\label{conj:deBcomp}
If $n \geq 3$ and $c$ is an integer such that $2^{n-1} +n \leq c \leq 2^n-1$ and $c \neq 2^{n-1}+n+1$, then
there exists a span $n$ de Bruijn sequence such that $c=c(\cS)$.
\end{conjecture}

\subsection{The $\bD$-morphism}
\label{sec:Doperator}

A key operator in the recursive constructions which follow is the $\bD$-morphism, originally defined by Lempel~\cite{Lem70}.
The operator $\bD$ is a 2-to-1 mapping from $\F_2^n$ on to $\F_2^{n-1}$, where $n \geq 2$.
For a binary word $(x_1,x_2,\ldots,x_n)$ we have
$$
\bD (x_1,x_2,\ldots,x_n) = (x_1 + x_2, x_2+x_3,\ldots, x_{n-1} + x_n )= (\bE +\bfone)(x_1,x_2,\ldots,x_n)  .
$$
By this definition we have that $\bD x = \bD y$ if and only if $y=x$ or $y=\bar{x}$.
The operator $\bD$ can be applied to cyclic sequences (cycles).
For a cyclic sequence $[x_1,x_2,\ldots,x_n]$ we have
$$
\bD [x_1,x_2,\ldots,x_n] = [x_1 + x_2, x_2+x_3,\ldots, x_{n-1} + x_n ,x_n +x_1 ]=(\bE +\bfone)[x_1,x_2,\ldots,x_n] .
$$
The operator $\bD$ has an inverse $\bD^{-1}$. For a sequence $\cS=[x_1,x_2,\ldots,x_n]$ of even weight,
$\bD^{-1} \cS$ is a set that contains two complementary sequences of length $n$,
$$
\bD_0^{-1} \cS = [0, x_1, x_1 + x_2 + x_3,\cdots, \sum_{i=1}^{n-1} x_i ]
$$
and
$$
\bD_1^{-1} \cS = [1, 1+x_1, 1+x_1 + x_2 + x_3,\cdots, 1+\sum_{i=1}^{n-1} x_i ].
$$
For a sequence $\cS=[x_1,x_2,\ldots,x_n]$ of odd weight, $\bD^{-1} \cS$ contains one self-dual sequence that can be written as
$$
\bD_0^{-1} \cS = [ 0 ,x_1, x_1+x_2,\ldots, \sum_{i=1}^{n-1} x_i ,
1 ,1+x_1, 1+x_1+x_2,\ldots, 1+\sum_{i=1}^{n-1} x_i ]
$$
or as
$$
\bD_1^{-1} \cS = [ 1 ,1+x_1, 1+x_1+x_2,\ldots, 1+\sum_{i=1}^{n-1} x_i ,
0 ,x_1, x_1+x_2,\ldots, \sum_{i=1}^{n-1} x_i ] .
$$

It is important to note that when $\bD$ or $\bD_1^{-1}$ are applied on cyclic sequences the outcome is a cyclic sequence.
The definition of $\bD$ implies that $\bD = \bE + \bfone$ and hence some of the results concerning the
shift operator~$\bE$ associated with sequences of length $2^n$ can be given in terms of the operator $\bD$, e.g.,
\begin{lemma}
\label{lem:DcompS}
For a nonzero sequence $\cS$ of length $2^n$ we have $c(\cS) = c(\bD \cS) +1$.
\end{lemma}

Finally, we can apply the inverse operator $\bD^{-1}$ on a set of sequences $\cT$ as follows:
$$
\bD^{-1} \cT \triangleq \{ \bD^{-1} \cS ~:~ \cS \in \cT \}
$$
\begin{lemma}
If $\cF$ is a factor in $G_n$, then $\bD^{-1} \cF$ is a factor in $G_{n+1}$.
\end{lemma}

The operator $\bD$ and its inverse $\bD^{-1}$ can be applied recursively on a sequence $\cS$
and its derived sequences. It is defined for $k \geq 2$ as follows.
$$
\bD^k \cS = (\bE + \bfone)^k \cS = (\bE + \bfone) (\bE + \bfone)^{k-1} \cS
$$
and
$$
\bD^{-k} \cS = \bD^{-1} (\bD^{-k+1})\cS
$$

The equality $\bD = \bE + \bfone$ and the properties of the operator $\bD$, its inverse $\bD^{-1}$, and the linear complexity
imply the following.
\begin{lemma}
\label{lem:DinOnce}
If $\cF$ is a set of $\Delta$ vertex-disjoint cycles of length (period) $2^k$, with linear complexity smaller than~$2^k$
(and hence of even weight), in $G_n$, then
$\cD^{-1} \cF$ contains $2 \Delta$ vertex-disjoint cycles of length $2^k$ in $G_{n+1}$.
\end{lemma}
\begin{lemma}
\label{lem:DinOnceOdd}
If $\cF$ is a set of $\Delta$ vertex-disjoint cycles of length $2^k$, with linear complexity $2^k$
(and hence of odd weight), in $G_n$, then
$\cD^{-1} \cF$ contains $\Delta$ vertex-disjoint self-dual cycles of length $2^{k+1}$ in $G_{n+1}$.
\end{lemma}

\subsection{Perfect factors}
\label{sec:perfect_factor}

The perfect factors of $G_n$, defined below, play pivotal role in this paper.
\begin{definition}
A {\bf \emph{perfect factor}}, \textup{PF}$_{p^\alpha}(n,k)$ in $G_{p^\alpha,n}$, where $p$ is a prime and $\alpha \geq 1$,
is a set of $p^{\alpha n-k}$ vertex-disjoint cycles of length $p^k$ in $G_{p^\alpha,n}$, where w.l.o.g. we assume that the alphabet
is $\Z_{p^\alpha}$. On each cycle, one vertex ($n$-tuple) is chosen arbitrarily to be the {\bf \emph{zero state}}.
When $p=2$ and $\alpha =1$ the perfect factor
will be denoted by PF$(n,k)$. Only the case $p=2$ will be considered here and $\alpha > 1$ will be considered only in
Construction~\ref{con:FF_DB} and Theorem~\ref{thm:MA_gen}.
The cycles (sequences) of the factor are numbered by $X_0,X_1,\ldots,X_{2^{n-k} -1}$.
\end{definition}

The following theorem on the existence of perfect factors was proved in~\cite{Etz88}.
\begin{theorem}
\label{thm:PF_exist}
A perfect factor \textup{PF}$(n,k)$ exists if and only if $k \leq n < 2^k$, i.e., $n < 2^k \leq 2^n$.
\end{theorem}

The theorem was extended for larger alphabet by Paterson~\cite{Pat95}. The following case of his result is required
for Construction~\ref{con:FF_DB} and Theorem~\ref{thm:MA_gen}.
\begin{theorem}
A perfect factor PF$_{2^\alpha}(n,k)$ exists if and only $n+1 \leq 2^k \leq 2^{\alpha n}$.
\end{theorem}

For a given set of parameters there could be many distinct perfect factors. We will be interested in
perfect factors in which the weights of all the sequences are either even or odd.
Both types of factors will be
important in our exposition and they will be discussed in Section~\ref{sec:analysis}.

One type of PF$(n,n-1)$, whose sequences are $\cS$ and $\bar{\cS}$, is of special interest.
The sequence which contains the all-zero $n$-tuple in such a factor will be called
a span $n$ {\bf \emph{half de Bruijn sequence}}. In such a sequence either $x$ or $\bar{x}$ is contained in $\cS$, for any given
$n$-tuple~$x$.

\section{Direct Constructions for de Bruijn Array Codes}
\label{sec:direct_construct}

In this section, we present direct constructions for de Bruijn array codes.
These constructions can be used to generate input to the recursive constructions presented in Section~\ref{sec:recursive_construct}
and to the merge-or-split method applied on some DBACs in Section~\ref{sec:join}.
The section consists of three subsection. In Section~\ref{sec:graph_represent} we present a two-dimensional generalization of the de Bruijn graph
which yields also a two-dimensional generalization of feedback shift registers. Based on this generalization
in Section~\ref{sec:2d_FSR} two direct constructions of DBACs are presented. Two other direct constructions for DBACs are presented in
Section~\ref{sec:other_direct}.

\subsection{Two-dimensional FSR and de Bruijn graph}
\label{sec:graph_represent}

One-dimensional sequences can be viewed as cycles in the de Bruijn graph. We aim also to view the two-dimensional arrays
constructed here as cycles in the family of graphs defined now.
The graph for the representation of the two-dimensional codes was defined in~\cite{Etz88}.
This graph is appropriate to codes in which all the columns of the matrices that form the codewords are
sequences of some given perfect factor. For this purpose,
let $\cF$ be a PF$(n,k)$ and define a graph $G_{\cF,m}$ whose vertices are $2^k \times m$ matrices.
These matrices are periodic (cyclic) only vertically. Each such matrix $X$ can be represented as
$$
X=(X_{i_0},\bE^{j_1} X_{i_1},\ldots, \bE^{j_{m-1}} X_{i_{m-1}}),
$$
where $X_{i_\ell}$, $0 \leq \ell \leq m-1$, is a cycle in $\cF$ represented as a column vector
and $\bE^{j_\ell} X_{i_\ell}$ is a cyclic shift of $X_{i_\ell}$ up by $j_\ell$ positions compared to its zero state,
i.e., it is the column vector of $\bE^{j_\ell} X_{i_\ell}$.
The first cycle $X_{i_0}$ is taken in its zero shift, i.e., it starts with its zero state.
An edge in the graph is a $2^k \times (m+1)$ matrix which is periodic only vertically. Such a matrix
$$
Y=(Y_{i_0},\bE^{j_1} Y_{i_1},\ldots, \bE^{j_{m-1}} Y_{i_{m-1}},\bE^{j_m} Y_{i_m} )
$$
is an edge from the vertex
$$
(Y_{i_0},\bE^{j_1} Y_{i_1},\bE^{j_2} Y_{i_2},\ldots, \bE^{j_{m-1}} Y_{i_{m-1}})
$$
to the vertex
$$
(Y_{i_1}, \bE^{j_2-j_1} Y_{i_2}, \ldots, \bE^{j_{m-1}-j_1} Y_{i_{m-1}},\bE^{j_m-j_1} Y_{i_m} ).
$$
This implies that the graph $G_{\cF,m}$ has $2^{nm-k}$ vertices and $2^{n(m+1)-k}$ edges.
%A cycle of length $\ell$, with no repeated vertices, in $G_{\cF,m}$
%can be represented by a vertically periodic $2^k \times (\ell +m-1)$ matrix
%in which each $n \times m$ matrix is contained exactly once. If the $2^k \times (m-1)$
%matrix obtained from the first $m-1$ columns of this matrix
%equals (without a cyclic shift) to the $2^k \times (m-1)$ matrix obtained from the last $m-1$ columns,
%then the cycle can be represented by a $2^k \times \ell$.

%Similarly, an Eulerian cycle in $G_{\cF,m}$ can be represented by a vertically periodic $2^k \times (2^{n(m+1)-k} +m)$ matrix.
%If the $2^k \times m$ matrix obtained from the first $m$ columns of this matrix
%equals the $2^k \times m$ matrix obtained from the last $m$ columns, then the cycle can be represented by a $2^k \times 2^{n(m+1)-k}$
%doubly-periodic matrix. This is satisfied when $m > 1$ or $n \neq k$ and the associated matrix
%forms a $(2^k,2^{n(m+1)-k};n,m+1)$-DBA.

%We are not interested here in a code with one codeword as these codes are just de Bruijn arrays and it was proved by
%Paterson~\cite{Pat94} that the necessary conditions of Theorem~\ref{lem:condPMC} are sufficient for this case.
%However, these codes can be used in our recursive constructions.

A $(2^k,2^t;n,m)$-DBAC, where the columns of the arrays are taken from a PF$(n,k)$, say $\cF$,
with $\Delta$~codewords can be generated from $G_{\cF,m}$.
A code in which all the columns of the codewords are from~$\cF$ can be viewed as a factor in $G_{\cF,m}$.
In this factor of $G_{\cF,m}$ the associated $2^k \times 2^t$ matrices must be doubly-periodic.
In other words, this factor in $G_{\cF,m}$ with $\Delta$ cycles of the same length $2^t$ forms a $(2^k,2^t;n,m)$-DBAC.
Any cycle of length $2^t$ in $G_{\cF,m}$ can be represented by a vertically periodic
(but not horizontally periodic) $2^k \times (2^t+m-1)$ matrix $A$.
If the last $m-1$ columns of $A$ are the same as the first $m-1$ column (with no vertical cyclic shift), then the cycle can be represented by
$2^k \times t$ doubly periodic matrix $\cC$ and this matrix can be a codeword in a $(2^k,2^t;n,m)$-DBAC.
It should be noted that there are $(2^k,2^t;n,m)$-DBACs whose columns do not form a perfect factor in $G_n$.
In all the direct constructions presented in this paper the columns of the arrays will be cycles from a perfect factor.
Moreover, the recursive constructions will make use of the cycles in a perfect factor to determine
when $\bD_0^{-1}$ is applied on columns of the array and when $\bD_1^{-1}$ is applied.
This view of DBACs can be considered as a two-dimensional generalization of perfect factors.
Moreover, the graph $G_{\cF,m}$ can be viewed as a two-dimensional generalization of the
de Bruijn graph and a factor in the graph as a state diagram for a two-dimensional generalization of nonsingular feedback shift registers.

The generalization for two-dimensional feedback shift register using a perfect factor PF$(n,k)$ is straightforward.
Suppose that perfect factor $\cF$ has $2^{n-k}$ cycles $X_0,X_1,\ldots, X_{2^{n-k}-1}$ of length $2^k$ and the register
has $m$ cells with variables $x_1, x_2, \ldots , x_m$. At each stage the cell $\ell$-th cell, $1 \leq \ell \leq m$,
stores a column vector $X_{i_\ell}$ of length~$2^k$ which is a sequence of
$\cF$ in some shift $\bE^{j_\ell}$ related to its zero state, where $j_1 = 0$, i.e., the sequence in $x_1$
is taken in its zero state. Using the feedback function
$x_{m+1}=f(x)=f(x_1,x_2,\ldots,x_m)$ we have
$$
x_{m+1} = f(X_{i_1}, \bE^{j_2} X_{i_2}, \ldots , \bE^{j_m} X_{i_m}),
$$
i.e., $x_{m+1}$ equals $\bE^{j_{m+1}} X_{i_{m+1}}$, i.e., for each $\ell$, $1 \leq \ell \leq m-1$, the $\ell$-th
cell receives its information from the $(\ell+1)$-th cell and hence it stores $\bE^{j_{\ell+1}-j_2} X_{i_{\ell+1}}$.
The shifts are taken modulo~$2^k$, the length of the cycles in $\cF$, and they are reduced by $j_2$ since in $x_1$ the
sequence should be taken in its zero state.
Finally $x_m$ will store $\bE^{j_{m+1}-j_2} X_{i_{m+1}}$. The state diagram is a subgraph of $G_{\cF,m}$ and for the
feedback function $f$ it has the edge
$$
(X_{i_1}, \bE^{j_2} X_{i_2}, \ldots , \bE^{j_m} X_{i_m}) \longrightarrow (X_{i_2}, \bE^{j_3-j_2} X_{i_3}, \ldots , \bE^{j_m-j_2} X_{i_m},\bE^{j_{m+1}-j_2} X_{i_{m+1}})~.
$$
Given this definition and the state diagram of the feedback function we also call the vertices of $G_{\cF,m}$ {\bf \emph{states}}.
As in the one-dimensional case we define a nonsingular two-dimensional feedback shift register to be a feedback shift register
whose state diagram contains only cycles. We can now develop a theory
for two-dimensional feedback shift registers analogous to the one-dimensional case. Such theory
is beyond the scope of the current paper, although some references relevant to the current work will be made.

The merge-or-split method is a feature that can be generalized from a factor in $G_n$ to a factor in $G_{\cF,m}$, i.e., from
one-dimensional feedback shift register to two-dimensional feedback shift registers, where $\cF$ is a factor
in $G_n$ with cycles of length $2^k$.
For simplicity we will consider factors in $G_{\cF,m}$ in which all the cycles have the same length $2^t$ and each
associated $2^k \times 2^t$ matrix form an $(2^k,2^t;n,m)$-DBAC, i.e., the $2^k \times 2^t$ matrix is doubly-periodic
(as $2^k \times (2^t+m-1)$ vertically-periodic array the first $m-1$ columns equal to the last $m-1$ columns).
A set of vertices in $G_{\cF,m}$ represented by $2^k \times m$ matrices is called a {\bf \emph{companion set}} if the
$2^k \times (m-1)$ matrices defined by first $m-1$ columns of the matrices in the set are all equal.
Two distinct vertices from this set will be called {\bf \emph{companion vertices}} (states) analogously to such pair of vertices in $G_n$.
Two cycles which contain a pair of companion vertices from such a set, one vertex on each cycle,
can be combined into one cycle when the merge-or-split method is applied to this pair of companion vertices in these two cycles.
The associated $2^k \times 2^{t+1}$ matrix obtained from this join is doubly-periodic since the sum of shifts in each
cycle is 0 modulo $2^k$ which implies that sum of shifts in the combined cycle is also 0 modulo $2^k$. The reason is that
the shifts in the two cycles are also the shifts of the combined cycle.

\subsection{Codes from two-dimensional FSRs}
\label{sec:2d_FSR}

The first two direct constructions are based on the existence of some perfect factors, and both can be
viewed as a two-dimensional generalization of nonsingular FSR.

\begin{construction}
\label{con:2D_CSR}
Let $\cF$ be a PF$(n,k)$, $m = 2^t-1$, $k \leq t$, $2 \leq k<n < 2^k$, and define the following two-dimensional FSR
\begin{equation}
\label{eq:2D_CSR}
f(X_{i_1},\bE^{j_2} X_{i_2},\ldots,\bE^{j_m}X_{i_m})=\bE^{j_{m+1}} X_{i_{m+1}},
\end{equation}
where
$$
X_{i_r} \in \cF, ~ 1 \leq r \leq m +1, ~\sum_{r=1}^{m+1} i_r \equiv 1 ~(\mmod ~ 2^{n-k}),~ \sum_{r=2}^{m+1} j_r \equiv 0 ~(\mmod ~ 2^k), ~
$$
$$
0 \leq j_r \leq 2^k-1, ~ 2 \leq r \leq m +1 \},
$$
where each $X_{i_r}$ is a column vector and its cyclic shift $\bE^{j_r}$ is relative to its zero state.
Let $\C$ be the code defined by the cycles of $f$.
\end{construction}

\begin{theorem}
\label{thm:PMC_odd}
The code $\C$ which contains the cycles of the two-dimensional FSR defined in Construction~\ref{con:2D_CSR}
%Let $\cF$ be a PF$(n,k)$, $m = 2^t-1$, $t \geq k$, and define the following set of matrices
%$$
%\C \triangleq \{ [X_{i_1},\bE^{j_2} X_{i_2},\ldots,\bE^{j_m}X_{i_m},\bE^{j_{m+1}} X_{i_{m+1}}] ~:~
%$$
%$$
%X_{i_r} \in \cF, ~ 1 \leq r \leq m +1, ~\sum_{r=1}^{m+1} i_r \equiv 1 ~(\mmod ~ 2^{n-k}),~ \sum_{r=2}^{m+1} j_r \equiv 0 ~(\mmod ~ 2^k), ~
%$$
%$$
%0 \leq j_r \leq 2^k-1, ~ 2 \leq r \leq m +1 \},
%$$
%where each $X_{i_r}$ is a column vector and its cyclic shifts is related to its zero state.
is a $(2^k,2^t;n,m)$-DBAC of size $2^{n m -k-t}$.
\end{theorem}
\begin{IEEEproof}
First, observe that the cycles of $\C$ in the graph $G_{\cF,m}$ are of length $2^t$.
(this can be observed by noting that the sum of the indices in a cycle $i_r$ is always congruent to $1 ~(\mmod ~ 2^{n-k})$ and
the sum of their shifts is always congruent to $0 ~(\mmod ~ 2^k)$ since $t \geq k$).

We start by showing that the horizontal period of each cycle defined in $\C$ is ${m +1 = 2^t}$ and not smaller.
If the period of some cycle $\cC$ in $\C$ is smaller than $2^t$, i.e., the associated matrix has a smaller horizontal periodicity, then
it should be a divisor of $2^t$, i.e., a power of 2. Assume to the contrary that such a matrix has a horizontal
period $2^s$ for some $s$, ${0 \leq s < t}$ which implies that it consists of $2^{t-s}$ disjoint $2^k \times 2^s$ equal
vertically periodic matrices.
Since $\sum_{r=1}^{2^t} i_r \equiv 1 ~(\mmod ~ 2^{n-k})$, it follows that $\sum_{r=1}^{2^t} i_r = \alpha 2^{n-k} +1$,
for some positive integer $\alpha$, and hence
$\sum_{r=1}^{2^s} i_r = \frac{\alpha 2^{n-k} +1}{2^{t-s}}$. But, $\frac{\alpha 2^{n-k} +1}{2^{t-s}}$
is not an integer, a contradiction. Therefore, all the cycles of~$\C$ have no smaller horizontal periodicity,
i.e., they form $2^k \times 2^t$ arrays with horizontal period $2^t$.

We continue by computing the size of the code $\C$. Each $X_{i_r}$, $1 \leq r \leq m$, is taken arbitrarily from~$\cF$
and hence it can be chosen in $2^{n-k}$ distinct ways. The sequence $X_{i_{m+1}}$ is determined
by the equation ${\sum_{r=1}^{m+1} i_r \equiv 1 ~(\mmod ~ 2^{n-k})}$. The first cycle $X_{i_1}$ is taken in its zero state,
while the next $m -1$ cycles can be taken in any of their $2^k$ possible cyclic shifts. The shift of the last
cycle $X_{i_{m+1}}$ is determined by the equation $\sum_{r=2}^{m+1} j_r \equiv 0 ~(\mmod ~ 2^k)$.
Hence, each of the middle $m -1$ cycles of $\cF$ can be chosen in $2^{n-k} 2^k=2^n$ distinct ways. The first cycle
can be chosen in $2^{n-k}$ distinct ways and the last cycle and its shift are determined by the first $m$ cycles and their shifts.
Therefore, there are altogether $2^{n(m-1)}2^{n-k}=2^{nm -k}$ distinct ways to choose all the cycles and their shifts.

We calculate the number of cycles that are counted more than once in this enumeration. Each constructed cycle~$\cC$ and its associated
codeword can start with any one of its $m +1 = 2^t$ columns.
Such a column, say the $r$-th column, will be taken in the shift where its zero state is the first $n$-tuple.
Each other column, say the $\ell$-th column is taken in shift $j_\ell - j_r$, i.e, $\bE^{j_\ell - j_r} X_{i_\ell}$
and hence, the associated matrix is shifted vertically by $i_r$ compared to $\cC$ (and horizontally by $r-1$).
Therefore, each codeword~$\cC \in \C$ is counted exactly $2^t$
times in this calculation. Thus, the size of $\C$ is $2^{nm -k}/2^t =2^{n m -k-t}$. Note, that the sum of all the shifts,
in this new arrangement of the first column, will remain congruent to 0 modulo $2^k$ since $t \geq k$ and hence the sum of the new shifts will
be changed by $j_r 2^t$ modulo $2^k$ which is congruent to 0~modulo~$2^k$,
where the $r$-th column of~$\cC$ is the new first column.

Therefore, the number of $n \times m$ windows in all the codewords of $\C$ is $2^{n m -k-t} \cdot 2^k \cdot 2^t=2^{n m}$.
Hence, to complete the proof it is sufficient to show that each $n \times m$ matrix appears as a window in a codeword of $\C$.
Consider such an $n \times m$ matrix $A$. The columns of $A$ are $n$-tuples in cycles of $\cF$ and hence these cycles can be arranged
in a codeword~$\cC \in \C$ with all the necessary shifts as required by $A$. This arrangements can be guaranteed by the definition
of the first $m =2^t -1$ columns in a codeword of $\C$
to form a window with the matrix $A$ in $\cC$, since there are no constraints on the shifts in the
construction for the $m-1$ columns which follows the first.

Thus, $\C$ is a $(2^k,2^t;n,m)$-DBAC of size $2^{n m -k-t}$.
\end{IEEEproof}

Note that in Construction~\ref{con:2D_CSR},
the requirement for $m +1$ being $2^t$ and for $\sum_{r=1}^{2^t} i_r$ to be $1 ~(\mmod ~ 2^{n-k})$ is
to avoid any smaller horizontal periodicity in the array. The requirement for $t \geq k$ is ensures that after a vertical cyclic
shift of the array the sum of the cyclic shifts of the columns will remain congruent to 0 modulo~$2^k$. This is
guaranteed since the shifts are taken modulo $2^k$ and $2^k$ divides~$2^t$.
%If we let $t < k$ in Construction~\ref{con:2D_CSR}, then there exists a horizontal shift of a $2^k \times 2^t$ matrix
%that will give a sum which is not congruent to 0 modulo $2^k$. This yields more codewords and repeated $n \times (2^t-1)$ windows.
Finally, two-dimensional FSR defined in this construction is a two-dimensional generalization
of the CSR$_{2^t-1}$. In the CSR$_{2^t-1}$ all the cycles have odd weight implying that all the cycles in the state diagram
are of period $2^t$.

\begin{construction}
\label{con:2D_CCR}
Let $\cF$ be a PF$(n,k)$, where $\cF$ does not contain self-dual sequences and
$\cS \in \cF$ if and only if $\bar{\cS} \in \cF$. For each $2 <k < n < 2^k$ and $t \geq 1$ define the following two-dimensional FSR
\begin{equation}
\label{eq:2D_CCR}
f(X_{i_1},\bE^{j_2} X_{i_2},\ldots,\bE^{j_{2^t}} X_{i_{2^t}})= \bar{X}_{i_1},
\end{equation}
where
$$
X_{i_r} \in \cF, ~ 1 \leq r \leq 2^t, ~ 0 \leq j_r \leq 2^k-1, ~ 2 \leq r \leq 2^t,
$$
and each $X_{i_r}$ is a column vector and its cyclic shift $\bE^{j_r}$ is relative to its zero state.
Let $\C$ be the code defined by the cycles of $f$.
\end{construction}

\begin{theorem}
\label{thm:PMC_SD}
The code $\C$ which contains the cycles of the two-dimensional FSR defined in Construction~\ref{con:2D_CCR}
%Let $\cF$ be a PF$(n,k)$, where $\cF$ does not contain self-dual sequences and
%$\cS \in \cF$ if and only if $\bar{\cS} \in \cF$. Define the following set of matrices
%$$
%\C \triangleq \{ [X_{i_1},\bE^{j_2} X_{i_2},\ldots,\bE^{j_{2^t}} X_{i_{2^t}},
%\bar{X}_{i_1},\bE^{j_2} \bar{X}_{i_2},\ldots,\bE^{j_{2^t}}\bar{X}_{i_{2^t}} ] ~:~
%$$
%$$
%X_{i_r} \in \cF, ~ 1 \leq r \leq 2^t, ~ 0 \leq j_r \leq 2^k-1, ~ 2 \leq r \leq 2^t \},
%$$
%where each $X_{i_1}$, $1 \leq r \leq 2^t$, is a column vector and its cyclic shift is relative to its zero state.
is a $(2^k,2^{t+1};n,2^t)$-DBAC of size $2^{n 2^t -k-t-1}$.
\end{theorem}
\begin{IEEEproof}
First, observe that the codewords of $\C$ are cycles of length $2^{t+1}$ in the graph $G_{\cF,m}$ associated with $2^k \times 2^{t+1}$ matrices.

We start by showing that the horizonal period of each codeword in $\C$ is $2^{t+1}$.
If the period of some codeword $\cC$ in $\C$ is smaller than $2^{t+1}$, then
it must be a divisor of $2^t$, i.e., a power of 2. However, this will imply that the sequence $X_{i_1}$ is some shift
of the sequence $\bar{X}_{i_1}$, i.e., this sequence is self-dual which is impossible by our choice of $\cF$.
Therefore, all the codewords of $\C$ have no smaller horizontal periodicity
and hence they form $2^k \times 2^{t+1}$ doubly-periodic arrays.

We continue by computing the size of the code $\C$. Each $X_{i_r}$, $1 \leq r \leq 2^t$, is taken arbitrarily from~$\cF$
and hence it can be chosen in $2^{n-k}$ distinct ways.
The first cycle $X_{i_1}$ is taken in its zero state,
while the next $2^t -1$ cycles can be taken in any of their $2^k$ cyclic shifts.
The last $2^t$ cycles and their shifts are determined by the first $2^t$ cycles, i.e., these are complement cycles with the same shifts.
Hence, each of the first $2^t$ cycles, except for $X_{i_1}$, has $2^{n-k} 2^k=2^n$ possible distinct choices.
Therefore, there are $2^{n(2^t-1)} \cdot 2^{n-k}=2^{n2^t-k}$ distinct choices for the codewords in $\C$ and their shifts.

We must calculate the number of codewords of $\C$ which are counted more than once in this enumeration.
Each constructed codeword~$\cC$ can start with
any of its $2^{t+1}$ columns. This column will be taken in the shift where its zero state is the first $n$-tuple.
The other columns are defined by the order of the columns in $\cC$ and their shifts are
taken exactly as in $\cC$ relative to the first state of the new first column. Therefore, each codeword~$\cC$ is counted exactly $2^{t+1}$
times in this calculation. Thus, the size of $\C$ is $2^{n2^t -k}/2^{t+1} =2^{n 2^t -k-t-1}$.

Therefore, the number of $n \times 2^t$ windows in the all the codewords of
$\C$ is $2^{n 2^t -k-t-1} \cdot 2^k \cdot 2^{t+1}=2^{n 2^t}$.
Hence, to complete the proof it is sufficient to show that each $n \times 2^t$ matrix appears as a window in a codeword of $\C$.
Consider such a matrix $A$. The columns of $A$ are $n$-tuples in cycles of $\cF$ and hence these cycles can be rearranged in a codeword~$\cC$
to form a window equals to the matrix $A$, since there are no constraints in the construction on the order of $2^t$ consecutive columns.

Thus,  $\C$ is a $(2^k,2^{t+1};n,2^t)$-DBAC of size $2^{n 2^t -k-t-1}$.
\end{IEEEproof}

Note that in Construction~\ref{con:2D_CCR}, the requirement that the length of codewords is $2^{t+1}$ as any other length
which is not a power of 2 implies smaller horizontal periodicity of some arrays.
The codeword $\cC \in \C$ contains a sequence of cycles from $\cF$ that together behave like a self-dual sequence
which cannot have smaller horizontally periodicity if its length is a power of 2.
In other words, the factor of $G_{\cF,2^t}$ defined in this construction is a two-dimensional generalization
of the CCR$_{2^t}$. In the state diagram of the CCR$_{2^t}$ all the cycles also have period $2^{t+1}$.
Moreover, the cycles of both the CCR$_{2^t}$ and the ones of Construction~\ref{con:2D_CCR} have that the elements in the $i$-th entry
is the complement of the element in the $(i+2^t)$-th entry.

\subsection{Other direct constructions}
\label{sec:other_direct}

The first direct construction can be used to construct $(2^k,2^t;n,1)$-DBACs for all possible parameters.
The key idea is to notice that in each such array the columns are sequences of length $2^k$ such that each $n$-tuple
is contained in exactly one of these columns. This implies that if the size of the code is $\Delta$,
then the union of the sequences in all the
columns of all the $\Delta$ arrays forms a perfect factor in $G_n$ and vice versa.
This perfect factor has $2^{n-k}$ cycles of length $2^k$, i.e.,
it is a PF$(n,k)$. Moreover, for this perfect factor, $2^t$ can be any divisor of $2^{n-k}$.
These observations imply the following construction and its associated corollaries.

\begin{construction}
\label{con:one_column}
Let $\cF$ be an PF$(n,k)$ and let $t$ be any integer such that $1 \leq t \leq n-k$.
Let $\C$ be the code obtained from $2^{n-k-t}$ $~~2^k \times 2^t$ arrays, where the columns of each array consists of $2^t$ distinct
sequences of $\cF$ in any chosen shift.
\end{construction}

\begin{corollary}
An $(2^k,2^t; n,1)$-DBAC, $1 \leq t \leq n-k$, of size $2^{n-k-t}$ exists if and only if there exists a PF$(n,k)$.
\end{corollary}

\begin{corollary}
\label{cor:PF_DBAC}
An $(2^k,2^t; n,1)$-DBAC, $1 \leq t \leq n-k$, of size $2^{n-k-t}$ exists if and only if $k \leq n < 2^k$,
i.e., $n < 2^k \leq 2^n$.
\end{corollary}

\begin{corollary}
When $t=0$, a $(2^k,2^t; n,1)$-DBAC is a perfect factor and when $t=n-k$, a $(2^k,2^t; n,1)$-DBAC is a de Bruijn array.
\end{corollary}

The last direct construction is a generalization of a technique presented by Ma~\cite{Ma84}, where in~\cite{Ma84}
one sequence over $\Z_{2^n}$ is used, while we use all the sequences of a perfect factor over $\Z_{2^n}$.

\begin{construction}
\label{con:FF_DB}
Let $\cF$ be a PF$_{2^n} (m,t)$, where $n \geq 2$, $3 \leq m+1 < 2^t \leq 2^{nm}$, and the sum of entries in each sequence of $\cF$
is congruent to 0~modulo~$2^n$. Let $\cS$ be a span $n$ de Bruijn sequence.
For any sequence ${\cQ=[q_0,q_1,\ldots,q_{2^t-1}]}$ of $\cF$ generate the following codeword in a code $\C$:
$$
[\cS, \bE^{q_0} \cS, \bE^{q_0+q_1} \cS , \bE^{q_0+q_1+q_2} \cS, \ldots , \bE^{\sum_{i=0}^{2^t-2} q_i} \cS].
$$
\end{construction}

\begin{theorem}
\label{thm:MA_gen}
Construction~\ref{con:FF_DB} yields a $(2^n,2^t;n,m+1)$-DBAC.
\end{theorem}
\begin{IEEEproof}
The generated codeword in $\C$ form a $2^n \times 2^t$ matrix and, since $\cS$ is a cyclic sequence
and $\sum_{i=0}^{2^t-1} q_i \equiv 0 ~ (\mmod ~ 2^n)$,
it follows that the matrix is doubly-periodic. The reason is that $\bE^{\sum_{i=0}^{2^t-1} q_i} \cS = \cS$,
$\bE^{\sum_{i=0}^{2^t-1} q_i+q_0} \cS = \bE^{q_0} \cS$, $\bE^{\sum_{i=0}^{2^t-1} q_i+q_0+q_1} \cS = \bE^{q_0+q_1} \cS$,
and so on, which implies that after the first $2^t$ columns we
continue the columns of the array (codeword) from its beginning.
The perfect factor $\cF$ has $2^{nm-t}$ sequences of length $2^t$ and each $m$-tuple is contained as a window
of length $m$ in exactly one of the sequences of $\cF$. From each sequence of $\cF$ we generate one codeword in~$\C$
and hence the total number of windows in the codewords of $\C$ is $2^{nm-t} \cdot 2^n \cdot 2^t = 2^{n(m+1)}$. Therefore, to prove that
$\C$ is a $(2^n,2^t;n,m+1)$-DBAC it is sufficient to show that each $n \times (m+1)$ matrix is a window
in one of the codewords of $\C$. Let $A$ be such an $n \times (m+1)$ matrix. Each column of $A$ is an $n$-tuple in $\cS$. Hence,
we have that $A$ is contained in the $2^n \times (m+1)$ vertically periodic array
$$
\cS ,\bE^{b_1} \cS ,\bE^{b_2} \cS ,\ldots , \bE^{b_m} \cS ,
$$
for some $m$-tuple $(b_1,b_2,\ldots,b_m)$ over $\Z_{2^n}$. Consider now another $m$-tuple $\beta$ with entries from $\Z_{2^n}$, where
$$
\beta=(b_1, b_2 -b_1,b_3-b_2,b_4 -b_3,\ldots,b_m -b_{m-1})=(p_j,p_{j+1},\ldots,p_{j+m-1})
$$
and $j+m-1$ is taken modulo $2^t$ since the sequence of $\cF$ are cyclic.
Since each $m$-tuple is contained in a window of length $m$ in one sequence of $\cF$, it follows that
$\beta$ is contained as such a window in such a sequence.
The associated codeword in $\C$ has the following $m+1$ successive columns
$$
\bE^\ell \cS, \bE^{\ell +p_j} \cS, \bE^{\ell +p_j +p_{j+1}} \cS,\ldots,\bE^{\ell +\sum_{i=0}^{m-1} p_{j+i}} \cS
=\bE^\ell \cS , \bE^{\ell+b_1} \cS , \bE^{\ell+b_2} \cS , \ldots , \bE^{\ell+b_m} \cS~,
$$
for some $\ell$.
These $m+1$ consecutive columns contain the matrix $A$ in $n$ consecutive rows. Thus, $\C$ is a $(2^n,2^t;n,m+1)$-DBAC.
\end{IEEEproof}

Applying Theorem~\ref{thm:MA_gen} requires the existence of PF$_{2^n} (m,t)$, in which the sum of
entries in each sequence is congruent to 0 modulo $2^n$. This is the case for the perfect factors
constructed in~\cite{Pat95}.

\section{Recursive Constructions for de Bruijn Array Codes}
\label{sec:recursive_construct}

For a recursive construction, we are given a $(2^k,2^t;n,m)$-DBAC $\C$ with parameters specified
in the construction, and a perfect factor $\cF$ with parameters which depend on the parameters of $\C$.
The sequences of $\cF$ are used as indicators associated with the columns of the codewords of~$\C$,
which we use to decide whether to apply $\bD^{-1}_0$ or to apply $\bD^{-1}_1$.
If the weight of the columns of the codewords in $\C$ is even, then applying $\bD^{-1}_1$ on these columns yields
codewords of the same vertical length, while if their weight is odd, then the vertical length is doubled.
Other distinction between the constructions is the horizontal length of the obtained codewords, it can remain the same
or increased. Each construction yield codewords with different sizes. Moreover, the number of constructed
codewords in the new code vary between the different constructions. This enables to generate codes
with various parameters starting with DBACs obtained by the direct constructions and applying various
recursive constructions iteratively. Some analysis for these applications will be discussed in Section~\ref{sec:analysis}.
All the recursive constructions
described here are generalizations and modifications of the two constructions presented by Fan, Fan, Ma, and Siu~\cite{FFMS85}
which are applied on a single array.

\begin{construction}
\label{con:DB_PMC_direct}
Let $\C$ be a $(2^k,2^m;n,m)$-DBAC of size $\Delta$, where $k,n,m \geq 2$, whose columns are sequences with even weight.
Let $A=[ \cA_1 \cA_2 ~ \cdots ~ \cA_{2^m}]$ be any codeword (a $2^k \times 2^m$ doubly-periodic array) of $\C$. Let
$$
\cS= [ s_1 , s_2, \ldots , s_{2^m} ]
$$
be a span $m$ de Bruijn sequence.

For the code $\C'$ define a codeword (array) $B$.
Let the $i$-th column of $B$ be $B_i=\bD^{-1}_{s_i} \cA_i$, where $1 \leq i \leq 2^m$.
Construct such a codeword for each of the $2^m$ distinct cyclic shifts of $\cS$,
where for each shift $\bE^{j-1} \cS=[ s_j , s_{j+1}, \ldots , s_{2^m},s_1,s_2,\ldots,s_{j-1} ]$ a codeword like $B$ is constructed
where $\bE^{j-1} \cS$ replaces $\cS$ in the construction of $B$.
\end{construction}

\begin{theorem}
\label{thm:DB_PMC_direct}
The code $\C'$ defined in Construction~\ref{con:DB_PMC_direct}
is an $(2^k,2^m;n+1,m)$-DBAC of size $2^m \Delta$.
\end{theorem}
\begin{IEEEproof}
From the discussion in Section~\ref{sec:Doperator}, by applying $\bD^{-1}$ on a column
from~$A$, we obtain two complementary sequences of the same period $2^k$.
Since $A$ is a $2^k \times 2^m$ doubly-periodic array, it follows that also $B$ is a $2^k \times 2^m$ doubly-periodic array.

Since the code~$\C$ is of size $\Delta$ and each array $A$ is used to construct $2^m$ new codewords
by using the $2^m$ cyclic shifts of $\cS$, it follows that
the new code $\C'$ has $2^m \Delta$ codewords. Each codeword in $\C'$ has $2^{k+m}$ distinct windows of size $(n+1) \times m$,
giving a total of $2^{k+m} 2^m \Delta$ windows of size $(n+1) \times m$ in all arrays of $\C$.
Since $2^{k+m} 2^m \Delta = 2^{nm} 2^m = 2^{(n+1)m}$
(note that $2^{k+m} \Delta =2^{nm}$ because $\C$ is a $(2^k,2^m;n,m)$-DBAC with $\Delta$ codewords), it follows that
to complete the proof it is sufficient to show that all the $(n+1) \times m$ windows in the codewords of $\C'$ are distinct.

Assume for the contrary that there exist two distinct $(n+1) \times m$ windows $X$ and $Y$ in codewords of~$\C'$,
$B_1$~and $B_2$ (not necessarily distinct)
obtained from codewords $A_1$ and $A_2$ of $\C$ (not necessarily distinct), respectively, for which $X=Y$
(if $A_1=A_2$ then $X$ and $Y$ are windows in distinct positions).
Let $\cX$ and~$\cY$ be the two $n \times m$ arrays obtained from $X$ and~$Y$, respectively, by applying $\bD$
to the columns of $X$ and $Y$, respectively (note that these columns are acyclic sequences).
Clearly, if $X=Y$, then $\cX=\cY$. Since $A_1$ and $A_2$ are codewords in the $(2^k,2^m;n,m)$-$\textup{DBAC}$ $\C$,
it follows that all the $n \times m$ windows in its codewords are distinct and hence $\cX=\cY$ implies that they are obtained
from the same position of a codeword $A$, i.e., $A_1=A_2$, using distinct shifts of $\cS$. Moreover, $\cX=\cY$ implies that they are in columns
of $B_1$ and $B_2$ projected from two sets of $m$ columns associated with the same $m$-tuple in~$\cS$.
But, this implies that $B_1=B_2$ and $\cS$ is taken with the same shift to obtain $A$
and hence $X$ and $Y$ are the same $(n+1) \times m$ window with the same position in $A$, a contradiction.
Thus, $\C'$ is an $(2^k,2^m;n+1,m)$-$\textup{DBAC}$ of size $2^m \Delta$.
\end{IEEEproof}

The next construction is a generalization of Construction~\ref{con:DB_PMC_direct}
and therefore its analysis should be similar.

\begin{construction}
\label{con:PF_PMC_direct}
Let $\C$ be a $(2^k,2^{m-\ell};n,m)$-DBAC of size $\Delta$, where $\ell \geq 0$, $m < 2^{m-\ell}$, whose columns are sequences
with even weight and $k,n,m \geq 2$. Let $\cF$ be a PF$(m,m-\ell)$ and
$A=[ \cA_1 \cA_2 ~ \cdots ~ \cA_{2^m}]$ be any codeword (a $2^k \times 2^m$ doubly-periodic array) of $\C$. Let
$$
\cS= [ s_1 , s_2, \ldots , s_{2^m} ]
$$
be any sequence of $\cF$.

For the code $\C'$ we define a new array $B$.
Let the $i$-th column of $B$ be $B_i=\bD^{-1}_{s_i} \cA_i$, where
$1 \leq i \leq 2^m$. The same method is applied with the $2^{m-\ell}$ distinct cyclic shifts of $\cS$
to obtain more codewords for $\C'$.
\end{construction}

\begin{theorem}
\label{thm:PF_PMC_direct}
The code $\C'$ defined in Construction~\ref{con:PF_PMC_direct}
is an $(2^k,2^{m-\ell};n+1,m)$-DBAC of size $2^m \Delta$.
\end{theorem}
\begin{IEEEproof}
Note first that by Theorem~\ref{thm:PF_exist} there exists a PF$(m,m-\ell)$.
The proof is very similar to the proof of Theorem~\ref{thm:DB_PMC_direct} and hence it is omitted.
\end{IEEEproof}

\begin{construction}
\label{con:DB_even}
Let $\C$ be a $(2^k,2^t;n,m)$-$\textup{DBAC}$ of size $\Delta$, where $k,n,t,m \geq 2$, whose columns are sequences
with even weight. Let $A$ be any codeword in $\C$ and define
$$
\cA \triangleq [ \overbrace{A ~ A ~ \cdots ~ A}^{2^m ~ \text{copies~of}~ A} ].
$$
Let
$$
\cS= [ s_0 , s_1, \ldots , s_{2^m -1} ]
$$
be a span $m$ de Bruijn sequence whose first $m$ bits are \emph{zeros}.
Define the sequence
$$
B \triangleq ( \overbrace{0,0, \ldots ,0}^{2^t ~ \text{zeroes}},\overbrace{s_1,s_2,\ldots , s_{2^m -1},\ldots, s_1,s_2,\ldots , s_{2^m -1}}^{2^t ~ \text{copies~of}~s_1,s_2,\ldots,s_{2^m -1}} )
$$
of length $2^t + (2^m-1)2^t=2^{m+t}$ and suppose $B = (b_1,b_2,\ldots,b_{2^{m+t}})$.
Define the following codeword~$\cB$ in a new code $\C'$:
let the $i$-th column of $\cB$ be $\cB_i=\bD^{-1}_{b_i} \cA_i$, where $\cA_i$ is the $i$-th column of $\cA$
and $1 \leq i \leq 2^{m+t}$.
\end{construction}

\begin{theorem}
\label{thm:DB_even}
The code $\C'$ defined in Construction~\ref{con:DB_even}
is a $(2^k,2^{m+t} ;n+1,m)$-DBAC of size $\Delta$.
\end{theorem}
\begin{IEEEproof}
It is an immediate observation that $\cA$ is an $2^k \times 2^{m+t}$ array. If we apply the operator $\bD^{-1}$ to a column
from $A$, the outcome is two complementary sequences of the same length $2^k$.
Note also that the sequence $[s_1, s_2, \ldots , s_{2^m -1} ]$
is the shortened de Bruijn sequence obtained from~$\cS$ (starts with $m-1$ \emph{zeros}).

The codeword $\cB$ is an $2^k \times 2^{m+t}$ doubly-periodic array and hence it has $2^{k+m+t}$ windows
of size $(n+1) \times m$. Since $2^{k+m} 2^t \Delta = 2^m 2^{nm} = 2^{(n+1)m}$ (since $2^k 2^t \Delta=2^{nm}$), it follows that
to complete the proof it is sufficient to show that each $(n+1) \times m$ binary array is contained at most once as
an $(n+1) \times m$ window in some~$\cB$.

Assume for the contrary that there exist two distinct $(n+1) \times m$ windows $\cX$ and $\cY$ in two codewords~$\cB_1$ and $\cB_2$ of $\C'$
(not necessary distinct) for which $\cX=\cY$.
Let $X$ and $Y$ be the $n \times m$ arrays obtained from $\cX$ and~$\cY$, respectively, by applying $\bD$
to the columns of $\cX$ and $\cY$, respectively (note that these columns are acyclic sequences).
Clearly, if $\cX=\cY$, then $X=Y$. Since $A$ is a codeword in an $(2^k,2^t;n,m)$-$\textup{DBAC}$,
it follows that all its $n \times m$ windows are distinct and hence $X=Y$ implies that they are obtained
from the same position of $A$ in different locations of $\cA$. Moreover, $X=Y$ implies that $\cX$ and $\cY$ are in columns
of~$\cB_1$ and $\cB_2$ projected from two sets of $m$ columns associated with the same $m$-tuple in~$B_1$ and $B_2$.
The fact that g.c.d.$(2^t,2^m-1)=1$ implies that the projection of two sets of $m$ columns on $\cA$ and the associated
two sets of $m$ entries in $B_1$ and $B_2$ (where the two sets
of columns from $\cA$ yield the same $2^k \times m$ window) cannot yield
the same pair of $m$-tuples from $B_1$ and $B_2$ and hence cannot yield the same pair of $(n+1) \times m$ sub-matrices
in $\cB_1$ and $\cB_2$, a contradiction.
Thus, $\C'$ is a $(2^k,2^{m+t};n+1,m)$-$\textup{DBAC}$ of size $\Delta$.
\end{IEEEproof}

\begin{construction}
\label{con:halfDB_even}
Let $\C$ be a $(2^k,2^t;n,m)$-$\textup{DBAC}$ of size $\Delta$, where $3 \leq m <2^t$, and $2 \leq k,n$,
whose columns are sequences with even weight. Let $A$ be any codeword in $\C$ and define
$$
\cA \triangleq [ \overbrace{A ~ A ~ \cdots ~ A}^{2^{m-1} ~ \text{copies~of}~ A} ].
$$
Let
$$
\cS= [ s_0 , s_1, \ldots , s_{2^{m-1} -1} ]
$$
be a span $m$ half de Bruijn sequence whose first $m$ bits are \emph{zeros}.
Define the sequence
$$
B \triangleq ( \overbrace{0,0, \ldots ,0}^{2^t ~ \text{zeroes}},\overbrace{s_1,s_2,\ldots , s_{2^{m-1} -1},\ldots, s_1,s_2,\ldots , s_{2^{m-1} -1}}^{2^t ~ \text{copies~of}~s_1,s_2,\ldots,s_{2^{m-1} -1}} )
$$
of length $2^t + (2^{m-1}-1)2^t=2^{m+t-1}$ and suppose $B = (b_1,b_2,\ldots,b_{2^{m+t-1}})$.
Define the following codeword~$\cB$ in a new code $\C'$:
let the $i$-th column of $\cB$ be $\cB_i=\bD^{-1}_{b_i} \cA_i$, where $\cA_i$ is the $i$-th column of~$\cA$
and $1 \leq i \leq 2^{m+t-1}$. Add the complement of $\cB$, $\bar{\cB}$, to $\C'$.
\end{construction}

The analysis for Construction~\ref{con:halfDB_even} is very similar to that for Construction~\ref{con:DB_even}
as well as the proof of the following theorem whose proof is the same as that for Theorem~\ref{thm:DB_even}.
\begin{theorem}
\label{thm:halfDB_even}
The code $\C'$ defined in Construction~\ref{con:halfDB_even}
exists a $(2^k,2^{m+t-1};n+1,m)$-DBAC of size $2\Delta$.
\end{theorem}

\begin{construction}
\label{con:DBodd_PMC_direct}
Let $\C$ be a $(2^k,2^t;n,m)$-DBAC of size $\Delta$, where $3 \leq m < 2^t$ and $2 \leq n < 2^k$,
whose columns are sequences with odd weight.
Let $A=[ \cA_1 \cA_2 ~ \cdots ~ \cA_{2^t}]$ be any codeword (a $2^k \times 2^t$ doubly-periodic array) of $\C$.
Let $\cF$ be a PF$(m,t)$ with no self-dual sequences,
where $\cC \in \cF$ if and only if $\bar{\cC} \notin \cF$. For each pair of sequences $\{\cC,\bar{\cC}\}$,
let $\cS=\cC$ or $\cS=\bar{\cC}$, where $\cC \in \cF$, and let
$$
\cS= [ s_1 , s_2, \ldots , s_{2^t} ].
$$
For the code $\C'$ define a new doubly-periodic array $B$.
Let the $i$-th column of $B$ be $B_i=\bD^{-1}_{s_i} \cA_i$, where
$1 \leq i \leq 2^t$. Apply the same approach on each of the $2^t$ distinct cyclic shifts of $\cS$ to obtain $2^t$ codewords in $\C'$.
\end{construction}

\begin{theorem}
\label{thm:DBodd_PMC_direct}
The code $\C'$ defined in Construction~\ref{con:DBodd_PMC_direct}
is a $(2^{k+1},2^t;n+1,m)$-DBAC of size $2^{m-1} \Delta$.
\end{theorem}
\begin{IEEEproof}
It is readily verified that a codeword $B$ in $\C'$ is a $2^{k+1} \times 2^t$ array.

The perfect factor $\cF$ has $2^{m-t}$ cycles and hence $2^{m-t-1}$ pairs of sequences $\{\cC,\bar{\cC}\}$.
This implies that each array $A$ is used to construct $2^{m-t-1} 2^t = 2^{m-1}$ codewords in $\C'$.
Since $\C$ is of size~$\Delta$ and each array $A$ is used to construct $2^{m-t-1} 2^t = 2^{m-1}$ codewords in $\C'$, it follows that
$\C'$ has $2^{m-1} \Delta$ codewords. Each codeword of $\C'$ has $2^{k+t+1}$ distinct windows of size $(n+1) \times m$
giving total of $2^{m-1} \Delta 2^{k+t+1}$ windows of size $(n+1) \times m$ in all the arrays.
Since $2^{m-1} \Delta 2^{k+t+1}= 2^{nm} 2^m = 2^{(n+1)m}$ (because $\Delta 2^{k+t}=2^{nm}$), it follows that
to complete the proof it is sufficient to show that each $(n+1) \times m$ array is contained at most once as
an $(n+1) \times m$ window in one codeword of $\C'$.

Suppose to the contrary that there exist two distinct $(n+1) \times m$ windows $X$ and $Y$ in codewords
of~$\C'$, $B_1$ and $B_2$ (not necessarily distinct)
obtained from codewords $A_1$ and $A_2$ of $\C$ (not necessarily distinct), respectively, for which $X=Y$.
Let $\cX$ and $\cY$ be the two $n \times m$ arrays obtained from $X$ and~$Y$, respectively, by applying $\bD$
to the columns of $X$ and $Y$, respectively (note that these columns are acyclic sequences).
Clearly, if $X=Y$, then $\cX=\cY$. Since $A_1$ and $A_2$ are codewords in an $(2^k,2^t;n,m)$-$\textup{DBAC}$,
it follows that all the $n \times m$ windows in all its codewords are distinct and hence $\cX=\cY$ implies that they are obtained
from the same position of a codeword $A$, i.e., $A_1=A_2$, using shifts of two sequences $\cS_1$ and $\cS_2$. Moreover, $\cX=\cY$ implies
that they are in columns of $B_1$ and $B_2$ projected from two sets of $m$~columns associated with the same $m$-tuple $\cS$
(either $\cS_1$ or $\cS_2$). But, this implies that $B_1=B_2$ and $\cS$ is taken with the same shift to obtain $A$
and hence $X$ and $Y$ are the same window with the same position in $A$, a contradiction.
Thus, $\C'$ is an $(2^{k+1},2^t;n+1,m)$-$\textup{DBAC}$ of size $2^{m-1} \Delta$.
\end{IEEEproof}

\begin{construction}
\label{con:lastODD}
Let $\C$ be a $(2^k,2^t;n,m)$-$\textup{DBAC}$ of size $\Delta$, where $3 \leq m < 2^t$ and $2 \leq n < 2^k$,
whose columns are sequences with odd weight. Let $A$ be any codeword in $\C$ and define
$$
\cA \triangleq [ \overbrace{A ~ A ~ \cdots ~ A}^{2^{m-1} ~ \text{copies~of}~ A} ].
$$
Let
$$
\cS= [ s_0 , s_1, \ldots , s_{2^{m-1} -1} ]
$$
be a span $m$ half de Bruijn sequence whose first $m$ bits are \emph{zeros} (therefore
$[s_1, s_2, \ldots , s_{2^{m-1} -1} ]$
is the shortened half de Bruijn sequence starting with $m-1$ \emph{zeros}). Define the sequence
$$
B \triangleq ( \overbrace{0,0, \ldots ,0}^{2^t ~ \text{copies}},\overbrace{s_1,s_2,\ldots , s_{2^{m-1} -1},\ldots, s_1,s_2,\ldots , s_{2^{m-1} -1}}^{2^t ~ \text{copies~of}~s_1,s_2,\ldots,s_{2^{m-1} -1}} )
$$
of length $2^t + (2^{m-1}-1)2^t=2^{m+t-1}$ and suppose $B = (b_1,b_2,\ldots,b_{2^{m+t-1} })$.

Define an array $\cB$ in a new code $\C'$.
Let the $i$-th column of $\cB$ be $\cB_i=\bD^{-1}_{b_i} \cA_i$, where $\cA_i$ is the $i$-th column of $\cA$
and $1 \leq i \leq 2^{m+t-1}$.
\end{construction}

\begin{theorem}
\label{thm:lastODD}
The code $\C'$ defined in Construction~\ref{con:lastODD}
is a $(2^{k+1},2^{m+t-1};n+1,m)$-DBAC of size $\Delta$.
\end{theorem}
\begin{IEEEproof}
A codeword $\cB$ in $\C'$ is a $2^{k+1} \times 2^{m+t-1}$ doubly-periodic array and hence it has $2^{k+m+t}$
windows of size $(n+1) \times m$. Since $\Delta 2^{k+m+t} = 2^{m+nm} = 2^{(n+1)m}$ (because $\Delta 2^{k+t}=2^{nm}$), it follows that
to complete the proof it is sufficient to show that the $(n+1) \times m$ windows in the codewords of $\C'$ are distinct.

%Assume for the contrary that there exist two $(n+1) \times m$ windows $\cX$ and $\cY$ in~$\cB_1$ and $\cB_2$
%(not necessary distinct) for which $\cX=\cY$.
%Let $X$ and $Y$ be the two $n \times m$ arrays obtained from $\cX$ and~$\cY$, respectively, by applying $\bD$
%on the columns of $\cX$ and $\cY$, respectively (note that these columns are acyclic sequences).
%Clearly, if $\cX=\cY$, then $X=Y$. Since $A$ is a codeword in an $(2^k,2^t;n,m)$-$\textup{DBAC}$,
%it follows that all its $n \times m$ windows are distinct and hence $X=Y$ implies that they are obtained
%from the same position of $A$ in different locations in $\cA$. Moreover, $X=Y$ implies that $\cX$ and $\cY$ are in columns
%of~$\cB_1$ and $\cB_2$ projected from two sets of $m$ columns associated with the same $m$-tuple in~$B_1$ and $B_2$.
%Since g.c.d.$(2^t,2^m-1)=1$. This implies that the projection of two sets of $m$ columns on $\cA$ and the associated
%two sets of $m$ entries in $B_1$ and $B_2$ (where the two sets
%of columns from $\cA$ yield the same $2^k \times m$ window) cannot yield
%the same pair of $m$-tuples from $B_1$ and $B_2$ and hence cannot yield the same pair of $(n+1) \times m$ sub-matrices
%in $\cB_1$ and $\cB_2$, a contradiction.

The rest is the proof is very similar to that for Theorem~\ref{thm:DB_even} and hence is omitted.
Thus, $\C'$ is an $(2^{k+1},2^{m+t-1};n+1,m)$-$\textup{DBAC}$ of size $\Delta$.
\end{IEEEproof}

Each recursive construction can be applied with the codes obtained in the direct constructions or iteratively ar with codes
obtained by other recursive construction. Also note that a $(2^k,2^t;m,n)$-DBAC is equivalent to a $(2^t,2^k;n,m)$-DBAC by transposing
the codewords of the code. The recursive constructions can used such a code if the obtained code satisfies the requirements
on the weight of the columns.

\section{Joining Cycles of $G_{\cF,m}$ }
\label{sec:join}

In this section, we provide a construction for $(2^k,2^t;n,m)$-DBACs from $(2^k,2^{t'};n,m)$-DBACs, where $t \geq  t'$.
This construction is based on joining cycles of the previously constructed DBACs, when the codewords are considered as
cycles in the graph $G_{\cF,m}$. The construction starts with
a $(2^k,2^t;n,m)$-DBAC of size $\Delta$, where all the columns of the codewords are from a perfect factor PF$(n,k)$ $\cF$,
and hence they are cycles in $G_{\cF,m}$. The codewords must be ordered in pairs
such that each pair of codewords contains a pair of companion vertices, i.e., if $\{\cC_1,\cC_2\}$ is such a pair of codewords, then
there are two companion vertices, one on $\cC_1$ and a second on $\cC_2$.
We next join each pair by their companion vertices using the merge-or-split method. The code obtained is
a $(2^k,2^{t+1};n,m)$-DBAC of size $\Delta /2$. This process can be continued and applied to the newly generated DBAC.
Note that if the sum of shifts in one codeword is congruent to 0 modulo $2^k$ and in the second codeword the sum
is also congruent to 0 modulo $2^k$, then in the merged cycles (codewords) the sum of shifts
is also congruent to 0 modulo $2^k$. Therefore, merging two cycles which are represented by two doubly-periodic arrays, using the merge-or-split
method, yields a new cycle which can be represented by a doubly-periodic array.

We demonstrate the idea with two examples based on codes obtained using Constructions~\ref{con:2D_CSR} and~\ref{con:2D_CCR}.
The first example is comprehensive while the second example is only partial.

The first example starts with the cycles of
a $(2^k,2^{t+1};n,2^t)$-DBAC $\C$ obtained from Construction~\ref{con:2D_CCR} with $\cF$ as the PF$(n,k)$,
where $\cF$ has no self-dual sequences and $\cS \in \cF$ if and only if $\bar{\cS} \in \cF$.
This code has $2^{n 2^t -k-t-1}$ codewords. Our goal is to order all the codewords of $\C$ in a cyclic list (as all the lists
in the two examples in this section) such that
each two consecutive codewords in this list have a pair of companion vertices and hence can be combined
using the merge-or-split method. Since this list has $2^{n 2^t -k-t-1}$ codewords, it will be possible to merge any sequence
of $2^s$ consecutive codewords, where $0 \leq s \leq  n 2^t -k-t-1$, and obtain a $(2^k,2^{t+1+s};n,2^t)$-DBAC.

We start with any ordering of the cycles of $\cF$
$$
X_0,X_1, \ldots , X_{2^{n-k-1}-1},X_{2^{n-k-1}},\ldots,X_{2^{n-k}-1},
$$
where $X_{2^{n-k-1}+i} =\bar{X}_i$ and $0 \leq i \leq 2^{n-k-1}-1$. We define the following simple mapping
$$
\varphi : \cF \longrightarrow \F_2
$$
where $\varphi (X_i) =0$ if $0 \leq i \leq 2^{n-k-1}-1$ and $\varphi (X_i) =1$ if $2^{n-k-1} \leq i \leq 2^{n-k}-1$.
The mapping $\varphi$ induces a mapping from the codewords of $\C$ into the set of self-dual sequences of length $2^{t+1}$.
Such a mapping maps each codeword of $\C$ into exactly one self-dual sequence of length $2^{t+1}$.
\begin{lemma}
\label{lem:mapped_zero}
The number of codewords of $\C$ which are mapped into a given self-dual sequence $\cS$ of length $2^{t+1}$ is $2^{(n-k-1)2^t}2^{k(2^t-1)}$.
\end{lemma}
\begin{IEEEproof}
Given a self-dual sequence $\cS$ of length $2^{t+1}$, it has exactly $2^t$ coordinates with \emph{zeros}. On each of these
coordinates we can choose any of the $2^{n-k-1}$ cycles from $\{ X_0,X_1, \ldots , X_{2^{n-k-1}-1} \}$ giving a total
of $2^{(n-k-1)2^t}$ distinct choices. Fixing any coordinate with a \emph{zero} and taking its cycle with a zero shift,
in each of the other $2^t -1$ coordinates with
\emph{zeros}, we can use any of the $2^k$ possible shifts of the cycle compared to its zero state giving a total of $2^{k(2^t-1)}$
distinct choices. Thus, the result follows.
\end{IEEEproof}

It was shown in~\cite{EtPa96} that the self-dual sequences of period $2^{t+1}$, $t \geq 3$, can be ordered
in a way such that each two consecutive sequences differ in exactly two positions $\ell$ and $\ell+2^t$. This ordering
will be very useful in obtaining the final list with all the codewords in $\C$. If $t=1$ then there is only one
self-dual sequence of length 4, $[0011]$. If $t=2$ there are two self-dual sequences of length 8 and their ordering is
$[00001111]$, $[00101101]$.

All the codewords which are mapped into the same self-dual
sequence are partitioned again into equivalence classes to simplify the understanding of the ordering.
Consider a self-dual sequence $\cS$ of length $2^{t+1}$ and its $2^t$ coordinates which are \emph{zeros}.
One of these coordinates is fixed and it induces an order on all the $2^t$ coordinates which are \emph{zeros}.
Using this order let
$$
(X_{i_1}, X_{i_2} ,\ldots,  X_{i_{2^t}})
$$
be the cycles of $\cF$ assigned to these $2^t$ coordinates in this order
such that $0 \leq i_j \leq 2^{n-k-1} -1$.
There are $2^{k (2^t-1)}$ codewords of this type for $\cS$ (having $(X_{i_1}, X_{i_2} ,\ldots,  X_{i_{2^t}})$ in these $2^t$ coordinates)
associated with the shifts of these cycles compared to their
zero states, where $X_{i_1}$ will be taken in its zero state.

The next step is to order these $2^{k (2^t-1)}$ codewords in such a way that each two consecutive codewords have a pair of companion vertices.
For this purpose we consider the set of ordered shifts of $(X_{i_1}, X_{i_2} ,\ldots,  X_{i_{2^t}})$, i.e., let
$$
\cQ_1 \triangleq \{ (0,j_2,j_3,\ldots,j_{2^t}) ~:~ 0 \leq j_\ell \leq 2^k -1, ~ 2 \leq \ell \leq 2^t \} .
$$
The pair $\{(0,j_2,j_3,\ldots,j_{2^t}),(X_{i_1}, X_{i_2} ,\ldots, X_{i_{2^t}})\}$
is associated with the following shifts of these cycles of $\cF$ in a codeword $\cC \in \C$,
$$
(X_{i_1}, \bE^{j_2} X_{i_2} ,\ldots, \bE^{j_{2^t}} X_{i_{2^t}}),
$$
where these entries (columns) are usually not consecutive in the codeword.
Hence, there exists a one-to-one mapping between the elements in the set $\cQ_1$ and the number of codewords represented by
$$
(X_{i_1}, X_{i_2} ,\ldots,  X_{i_{2^t}})
$$
for the self-dual sequence $\cS$.

Our next step in combining the codewords is to order the $2^{k(2^t-1)}$ elements of $\cQ_1$ in a cyclic Gray code, i.e., so that each
two elements will differ in exactly one position. Such a cyclic Gray code is well-known~\cite{Mut23} and is also very easy to construct.
The following lemma can be readily verified.
\begin{lemma}
\label{lem:adjacent_codewordsCCR}
If $(0,j_2,j_3,\ldots,j_{2^t}) \in \cQ_1$ and $(0,j'_2,j'_3,\ldots,j'_{2^t}) \in \cQ_1$
differ in exactly one coordinate, then their associated codewords in the same equivalence class
$$
(X_{i_1}, \bE^{j_2} X_{i_2} ,\ldots, \bE^{j_{2^t}} X_{i_{2^t}}),
$$
and
$$
(X_{i_1}, \bE^{j'_2} X_{i_2} ,\ldots, \bE^{j'_{2^t}} X_{i_{2^t}}).
$$
can be combined into one codeword using the merge-or-split method, i.e., they have a pair of companion vertices.
\end{lemma}
\begin{IEEEproof}
Since the two codewords in $Q_1$ differ in exactly one coordinate, the associated two codewords of $\C$, $\cC_1$ and $\cC_2$, differ
in exactly two coordinates separated by $2^t-1$ coordinates in which both $\cC_1$ and $\cC_2$ are identical.
The two $2^k \times 2^t$ matrices of $\cC_1$ and $\cC_2$
associated with these $2^t-1$ coordinates and the coordinate which follows them differ only in the shift of the last column.
This implies that these two $2^k \times 2^t$ matrices of $\cC_1$ and $\cC_2$ are companion vertices.
\end{IEEEproof}
The existence of a cyclic Gray code of the elements of the set $\cQ_1$ and Lemma~\ref{lem:adjacent_codewordsCCR} mean that we can form
a cyclic Gray code from the codewords of each equivalence class.

The next step is to order the codewords from different equivalence classes associated
with the same self-dual sequence $\cS$. For this purpose we define another Gray code on the set of elements
$$
\cQ_2 \triangleq \{ (i_1,i_2,i_3,\ldots,i_{2^t}) ~:~ 0 \leq i_\ell \leq 2^{n-k-1} -1, ~ 1 \leq \ell \leq 2^t \} .
$$
The element $(i_1,i_2,i_3,\ldots,i_{2^t}) \in Q_2$ is associated with the equivalence class
$$
(X_{i_1}, X_{i_2} ,\ldots,  X_{i_{2^t}}).
$$
It is again easy to verify that we can find a pair of companion vertices in two distinct equivalence classes
associated with the same self-dual sequence $\cS$, which differ in exactly one coordinate of $Q_2$.
\begin{lemma}
\label{lem:adjacent_Q2}
If $(i_i,i_2,i_3,\ldots,i_{2^t}) \in \cQ_2$ and $(i'_1,i'_2,i'_3,\ldots,i'_{2^t}) \in \cQ_1$
differ in exactly one coordinate, then their associated codewords (where
all the cycles are the same and are also in the same shift) in the associated equivalence classes
$$
(X_{i_1}, X_{i_2} ,\ldots, X_{i_{2^t}}),
$$
and
$$
(X_{i'_1}, X_{i'_2} ,\ldots, X_{i'_{2^t}}),
$$
can be combined into one codeword using the merge-or-split method, i.e., they have a pair of companion vertices.
\end{lemma}
\begin{IEEEproof}
These two codewords differs in two coordinates as in Lemma~\ref{lem:adjacent_codewordsCCR},
and hence have a pair of companion vertices.
\end{IEEEproof}

%Lemmas~\ref{lem:adjacent_codewordsCCR} and~\ref{lem:adjacent_Q2} imply that we can order
%all the codewords associated with the same self-dual sequence $\cS$ in a cyclic list such that each two
%consecutive elements in the list have a pair of companion vertices. Hence, we have the following lemma.
%\begin{lemma}
%All the codewords associated with the same self-dual sequence $\cS$ can be ordered in a cyclic list in which any consecutive
%codewords have a pair of companion vertices.
%\end{lemma}
We next consider two self-dual sequences of length $2^{t+1}$, $\cS_1$ and $\cS_2$ which differ in exactly
two coordinates $i$ and $i+2^t$.

\begin{lemma}
\label{lem:adjacent_SD_C}
Any two self-dual sequences of length $2^{t+1}$, which differ in exactly two coordinates have in their associated codewords
a pair of codewords whose vertices have a pair of companion vertices.
\end{lemma}
\begin{IEEEproof}
As before we can use two associated codewords which differ in their cycles only in these two coordinates. Again, we can use
all the cycles in their zero states and hence the two associated codewords will have a pair of companion vertices.
\end{IEEEproof}

The last step is to use Lemmas~\ref{lem:adjacent_codewordsCCR},~\ref{lem:adjacent_Q2}, and~\ref{lem:adjacent_SD_C}
to combine all the cyclic lists into one cyclic list which contains all the codewords,
where each two consecutive codewords have a pair of companion vertices.
\begin{theorem}
All the codewords of a $(2^k,2^{t+1};n,2^t)$-DBAC $\C$ obtained in Construction~\ref{con:2D_CCR} can be ordered
in a cyclic list in which each two consecutive codewords have a pair of companion vertices.
\end{theorem}
\begin{IEEEproof}
We start by merging together two cyclic lists associated with the same self-dual sequence $\cS$ and two different equivalence classes,
which are the first two in the list of equivalence classes associated with $Q_2$,
$$
(X_{i_1}, X_{i_2} ,\ldots, X_{i_{2^t}}),
$$
and
$$
(X_{i'_1}, X_{i'_2} ,\ldots, X_{i'_{2^t}}),
$$
which differ in exactly one coordinate. Assume further that the first element in the Gray code list of $Q_1$ is
$$
(0,j_2,j_3,\ldots,j_{2^t})
$$
and the last element is
$$
(0,j'_2,j'_3,\ldots,j'_{2^t}) ~.
$$

The combined list starts with
$$
(X_{i_1}, \bE^{j_2} X_{i_2} ,\ldots, \bE^{j_{2^t}} X_{i_{2^t}}),
$$
and continues with the elements of the same equivalence class until it reaches
$$
(X_{i_1}, \bE^{j'_2} X_{i_2} ,\ldots, \bE^{j'_{2^t}} X_{i_{2^t}}),
$$
The next element in the combined list is
$$
(X_{i'_1}, \bE^{j'_2} X_{i'_2} ,\ldots, \bE^{j'_{2^t}} X_{i'_{2^t}}).
$$
These last two elements differ in exactly one coordinate and hence they are associated with a pair of companion vertices.
We continue with this new equivalence class in reverse order until we reach the element
$$
(X_{i'_1}, \bE^{j_2} X_{i'_2} ,\ldots, \bE^{j_{2^t}} X_{i'_{2^t}}),
$$
which differs in exactly one coordinate from the first element
$$
(X_{i_1}, \bE^{j'_2} X_{i_2} ,\ldots, \bE^{j'_{2^t}} X_{i_{2^t}}).
$$
We continue in the same manner with the third equivalence class in the list in its usual order and the fourth equivalence
class in reverse order and so on until all the equivalence classes associated with $\cS$ are combined.

Once all the codewords of the equivalence classes for each self-dual sequence have been merged into one cyclic
list in which each two consecutive codewords have a pair of companion vertices, we need to combine the different
codewords of distinct self-dual sequences into one long cyclic list in which each two consecutive codewords have a pair of companion vertices.
Let $\cS_1$ and $\cS_2$ be the first two self-dual sequences in the list. By definition, $\cS_1$ and $\cS_2$ differ in
exactly two coordinates $\ell$ and $\ell +2^t$. For these two self-dual sequences, we take two equivalence classes
that differ only in the coordinates in which $\cS_1$ and $\cS_2$ differ and merge the two lists using the same approach as
the one used in merging the lists of two different equivalence classes for the same self-dual sequence.

The outcome is a cyclic list including all the codewords of $\C$, in which each two consecutive codewords have a pair
of companion vertices.
\end{IEEEproof}

\begin{corollary}
\label{cor:join_SD}
If $t$, $k$, $\ell$, and $n$ are integers such that $k < n < 2^k$, $1 \leq \ell$, and $\ell +1 \leq t \leq n 2^\ell -k$, then
there exists a $(2^k,2^t;n,2^\ell)$-DBAC of size $2^{n 2^\ell -k-t}$.
\end{corollary}

A nontrivial example ($t > 1$) for the list of codewords of a $(2^k,2^{t+1};n,2^t)$-DBAC with the property that each two codewords have
a pair of companion vertices is too large to include here and hence we omit such an example and leave it for the interested reader.
%Finally, $(2^k,2^{t+1+s};n,2^t)$-DBAC (as defined in Corollary~\ref{cor:join_SD}) can be also defined directly.

\vspace{0.2cm}

A second example for the idea of joining cycles in $G_{\cF,m}$
starts with the codewords of a $(2^k,2^t;n,2^t -1)$-DBAC $\C$ obtained from
Construction~\ref{con:2D_CSR} with $\cF$ as the PF$(n,k)$ which are ordered arbitrarily by
$X_0, X_1,\ldots,X_{2^{n-k}-1}$. Such a code has $2^{n m -k-t}$ codewords,
where $m = 2^t-1$ and $t \geq k$. Partition these codewords into equivalence classes (disjoint sets) as follows.
Each codeword $\cC \in \C$ is represented as a word of length $2^t$ as follows:
$$
(X_{i_1}, \bE^{j_2} X_{i_2} ,\ldots, \bE^{j_{2^t}} X_{i_{2^t}}),
$$
where this representation is the one among the $2^t$ cyclic shifts of $\cC$ with the smallest lexicographic value
for $(i_1, i_2 ,\ldots, i_{2^t})$ (although this lexicographic value has no influence on the construction).
The codeword $\cC$ is in the equivalence class
$$
(i_1, i_2 ,\ldots, i_{2^t}),
$$
Each $X_{i_j}$, except for the last, can be chosen in $2^{n-k}$ distinct ways for a total of $2^{(n-k) (2^t-1)}$ distinct
choices. The last is determined by the equation $\sum_{r=1}^{2^t} i_r \equiv 1 ~(\mmod ~ 2^{n-k})$.
Since each of the $2^t$ cycles from $\cF$ in the equivalence class
$$
(i_1, i_2 ,\ldots, i_{2^t})
$$
can be chosen as the first, it follows that the number of equivalence classes in this equivalence relation is
$$
2^{(n-k) (2^t-1)}/2^t = 2^{(n-k) (2^t-1)-t} ~.
$$
Merging the codewords in each equivalence class can be achieved exactly as in the first example. The first sequence is taken
in its zero shift. The next $2^t-2$ sequences taken in any of the $2^k$~distinct cyclic shifts. The shift of the
last sequence is chosen so that the sum of all the shifts is congruent to 0~modulo~$2^k$.
Two codewords of an equivalence class which differ in exactly two consecutive shifts have a pair of companion vertices since
the number of columns of the matrix associated with the vertex
is $2^t-1$ and these two codewords can be horizontally shifted in a way that they share
the first $2^t-2$ columns and their shifts. For this purpose we consider the set $Q_3$, whose elements define the shifts
of the sequences from $\cF$, is defined as follows.
$$
\cQ_3 \triangleq \{ (0,j_2,j_3,\ldots,j_{2^t}) ~:~ 0 \leq j_\ell \leq 2^k -1, ~ 2 \leq \ell \leq 2^t, ~ \sum_{\ell=2}^{2^t} j_\ell \equiv 0 ~ (\mmod ~ 2^k) \} ,
$$
where the element $(0,j_2,j_3,\ldots,j_{2^t}) \in Q_3$ for the equivalence class $(i_1, i_2 ,\ldots, i_{2^t})$ is associated with
the codeword
$$
(X_{i_1}, \bE^{j_2} X_{i_2} ,\ldots, \bE^{j_{2^t}} X_{i_{2^t}}).
$$
Order the elements of $Q_3$ in a cyclic list such that each pair of consecutive elements in the list differ in exactly two consecutive
elements. Such a list implies that each number of consecutive elements in the list can be combined via the merge-or-split method.
To have a longer cyclic list which includes all the codewords of $\C$ we need to order the equivalence classes in another cyclic list, where
each two consecutive equivalence classes in the list differ in exactly two consecutive elements.
Note, that these equivalence classes are over an alphabet of size $2^{n-k}$, the number of sequences in $\cF$. Two codewords
from consecutive elements in such a list which are associated with the same shifts from $Q_3$ differ in exactly
two coordinates and hence have a pair of companion vertices.

We do not have a general construction for such a list and we
leave it as a problem for future research. However, when the number of sequences in $\cF$ is two, i.e., $n-k=1$, it is relatively easy to
generate such a list. The list consists of all the cyclic sequences of length $2^t$ and odd weight.
Order these elements so that each two consecutive elements differ in exactly two coordinates.
For this purpose we consider the list of self-dual sequences of length $2^{t+1}$. If $Y_i$ and $Y_{i+1}$ are two consecutive
elements in this list, then $Y_i$ and $Y_{i+1}$ differ in two positions separated by $2^t-1$ other positions.
$\bD Y_i$ and $\bD Y_{i+1}$ have odd weight and they differ in exactly two adjacent positions. Thus, the existence
of a list for the self-dual sequences of length $2^{t+1}$ implies the list of sequences of odd weight and length $2^t$.

\begin{example}
If $n-k =1$, then there are two sequences in a PF$(n,k)$: $[0001]$ and $[0111]$, and these two sequences are the columns of all
the codewords in $\C$. When $t=2$ the equivalence classes are $(0,0,0,1)$ and $(0,1,1,1)$. The ordered list of these two equivalence classes
is trivial. A list with all the elements of $Q_3$ is given as follows (the elements of $Q_3$ are presented as column vectors).
$$
\begin{array}{c}
0000000000000000 \\
0110033332211022 \\
0301301323121202 \\
0033110233112200
\end{array}
$$
The code $\C$ has 32 codewords. The 16 codewords of the equivalence class $(0,0,0,1)$ are ordered in the list as follows.
$$
\left[
\begin{array}{c}
0 0 0 0 \\
0 0 0 1  \\
0 0 0 1  \\
1 1 1 1
\end{array}
\right] , ~
\left[
\begin{array}{c}
0 0 1 0 \\
0 0 0 1 \\
0 1 0 1 \\
1 0 0 1
\end{array}
\right] , ~
\left[
\begin{array}{c}
0 0 0 1 \\
0 0 0 0  \\
0 1 0 1  \\
1 0 1 1
\end{array}
\right] , ~
\left[
\begin{array}{c}
0 0 0 1 \\
0 0 0 0 \\
0 0 1 1 \\
1 1 0 1
\end{array}
\right] , ~
\left[
\begin{array}{c}
0 0 1 1 \\
0 0 0 1  \\
0 0 0 1  \\
1 1 0 0
\end{array}
\right] ,
\left[
\begin{array}{c}
0 1 0 1 \\
0 0 0 1 \\
0 0 0 1 \\
1 0 1 0
\end{array}
\right] , ~
\left[
\begin{array}{c}
0 1 0 0 \\
0 0 0 1  \\
0 0 1 1  \\
1 0 0 1
\end{array}
\right] , ~
\left[
\begin{array}{c}
0 1 1 1 \\
0 0 0 1 \\
0 0 0 0 \\
1 0 0 1
\end{array}
\right] ,
$$
$$
\left[
\begin{array}{c}
0 1 0 1 \\
0 0 1 0  \\
0 0 0 1  \\
1 0 0 1
\end{array}
\right] , ~
\left[
\begin{array}{c}
0 0 1 1 \\
0 1 0 0 \\
0 0 0 1 \\
1 0 0 1
\end{array}
\right] , ~
\left[
\begin{array}{c}
0 0 0 1 \\
0 1 0 1  \\
0 0 1 1  \\
1 0 0 0
\end{array}
\right] , ~
\left[
\begin{array}{c}
0 0 0 1 \\
0 0 1 1 \\
0 1 0 1 \\
1 0 0 0
\end{array}
\right] , ~
\left[
\begin{array}{c}
0 0 0 1  \\
0 0 0 1  \\
0 1 1 0  \\
1 0 0 1
\end{array}
\right] ,
\left[
\begin{array}{c}
0 0 0 1 \\
0 0 1 1 \\
0 0 0 0 \\
1 1 0 1
\end{array}
\right] , ~
\left[
\begin{array}{c}
0 0 0 1 \\
0 1 0 1  \\
0 0 0 0  \\
1 0 1 1
\end{array}
\right] , ~
\left[
\begin{array}{c}
0 0 0 0 \\
0 1 1 1 \\
0 0 0 1 \\
1 0 0 1
\end{array}
\right] ,
$$
The 16 codewords of the equivalence class $(0,1,1,1)$ are ordered in the list as follows.
$$
\left[
\begin{array}{c}
0 0 0 0 \\
0 1 1 1  \\
0 1 1 1  \\
1 1 1 1
\end{array}
\right] , ~
\left[
\begin{array}{c}
0 1 1 0 \\
0 1 0 1 \\
0 1 1 1 \\
1 0 1 1
\end{array}
\right] , ~
\left[
\begin{array}{c}
0 1 0 1 \\
0 1 1 0  \\
0 1 1 1  \\
1 0 1 1
\end{array}
\right] , ~
\left[
\begin{array}{c}
0 0 1 1 \\
0 1 1 0 \\
0 1 1 1 \\
1 1 0 1
\end{array}
\right] , ~
\left[
\begin{array}{c}
0 0 1 1 \\
0 1 0 1  \\
0 1 1 1  \\
1 1 1 0
\end{array}
\right] , ~
\left[
\begin{array}{c}
0 1 0 1 \\
0 0 1 1 \\
0 1 1 1 \\
1 1 1 0
\end{array}
\right] , ~
\left[
\begin{array}{c}
0 1 1 0 \\
0 0 1 1  \\
0 1 1 1  \\
1 1 0 1
\end{array}
\right] , ~
\left[
\begin{array}{c}
0 1 1 1 \\
0 0 0 1 \\
0 1 1 0 \\
1 1 1 1
\end{array}
\right] ,
$$
$$
\left[
\begin{array}{c}
0 1 1 1 \\
0 0 1 0  \\
0 1 0 1  \\
1 1 1 1
\end{array}
\right] , ~
\left[
\begin{array}{c}
0 1 1 1 \\
0 1 0 0 \\
0 0 1 1 \\
1 1 1 1
\end{array}
\right] , ~
\left[
\begin{array}{c}
0 1 1 1 \\
0 1 1 1  \\
0 0 1 1  \\
1 1 0 0
\end{array}
\right] , ~
\left[
\begin{array}{c}
0 1 1 1 \\
0 1 1 1 \\
0 1 0 1 \\
1 0 1 0
\end{array}
\right] , ~
\left[
\begin{array}{c}
0 1 1 1 \\
0 1 1 1  \\
0 1 1 0  \\
1 0 0 1
\end{array}
\right] , ~
\left[
\begin{array}{c}
0 0 1 1 \\
0 1 1 1 \\
0 1 0 0 \\
1 1 1 1
\end{array}
\right] , ~
\left[
\begin{array}{c}
0 1 0 1 \\
0 1 1 1 \\
0 0 1 0 \\
1 1 1 1
\end{array}
\right] , ~
\left[
\begin{array}{c}
0 1 1 0 \\
0 1 1 1 \\
0 0 0 1 \\
1 1 1 1
\end{array}
\right] .
$$
It is easy to verify that any set of consecutive codewords can be combined. Hence, we obtain
a $(4,2^{2+i};3,3)$-DBAC for each $i$, $0 \leq i \leq 5$.

\hfill\quad $\blacksquare $
\end{example}

\section{Analysis of the Constructions}
\label{sec:analysis}

In this section we summarize the parameters of DBACs which are obtained by our direct and recursive constructions including
the joining of cycles via the merge-or-split method.
We further apply the recursive constructions to characterize some families of DBACs which were completely solved
by our constructions.

We start with the existence of perfect factors which play an important role in our constructions.
Their weight of their sequences is also have an important role in the constructions. We can analyze their existence with either sequences
of odd weight or even weight. For sequences of even weight we can characterize all the parameters of perfect factors.

\begin{theorem}
\label{thm:FactorEven}
$~$
\begin{enumerate}
\item If $k \leq n < 2^k -1$, then there exists a PF$(n,k)$ for which all the sequences are of even weight.

\item If $n=2^k-1$ then there exists a unique PF$(n,k)$ and all the sequences in this factor are of odd weight.
\end{enumerate}
\end{theorem}
\begin{IEEEproof}
$~$
\begin{enumerate}
\item If $2^{k-1} \leq n < 2^k -1$, then by Lemma~\ref{lem:setsFor PF}
all the sequences of length $2^k$ which are generated by the polynomial $(\bE+\bfone)^{n+1}$
and are not generated by the polynomial $(\bE+\bfone)^n$ form a PF$(n,k)$.
By Lemma~\ref{lem:setsFor PF} all these sequences have linear complexity $n+1$, where $2^{k-1} < n+1 < 2^k$ and
hence by Lemma~\ref{lem:oddW_comp} all these sequences have even weight.

If $k \leq n < 2^{k-1}$, then consider a span $k$ de Bruijn sequence $\cS$ with linear complexity $2^{k-1}+k$
whose existence is implied by Theorem~\ref{thm:deBcomp}. By Lemmas~\ref{lem:DcompS},~\ref{lem:DinOnce}, and~\ref{lem:DinOnceOdd}
we have that $\bD^{-r} \cS$ for any $r$ such that $0 \leq r < 2^{k-1} -k$
contains $2^r$ sequences of length $2^k$
in $G_{k+r}$, i.e., $2^{n-k}$ sequences of length $2^k$ in $G_n$ which is a PF$(n,k)$ for any given $n$ such that $k \leq n < 2^{k-1}$.
By Lemma~\ref{lem:DcompS} all the generated sequences have linear complexity $2^{k-1} +k+r < 2^k$ and
hence by Lemma~\ref{lem:oddW_comp} all these sequences have even weight.

\item For $n=2^k-1$ let $\cF$ be a perfect factor PF$(n,k)$ with sequences of length $n+1=2^k$. One of these sequences
contains the all-zero word of length~$n$.
Clearly, the sequence which contains the all-zero word of length $n+1$ is $[0^n 1]$. This sequence contains all words of length $n$ and
weight one. Since each cyclic sequence of length $n+1$ and weight two contains exactly two words of weight one (by omitting a \emph{one}
from the sequence), it follows that $\cF$ does not contain sequences of weight two and hence all words
of weight two are contained in sequences with weight three. This implies that $\cF$ does not contain sequences of weight four and so on
by induction. Thus, there is a unique factor whose sequences have odd weight. This factor is the state diagram of the CSR$_n$.
\end{enumerate}
\end{IEEEproof}

Theorem~\ref{thm:FactorEven} implies that whenever a perfect factor PF$(n,k)$ exists, except for $n=2^k-1$, i.e., when $k \leq n < 2^k -1$ by
Theorem~\ref{thm:PF_exist}, there exists a perfect factor for which all its sequences have even weight. The situation is
more complicated for odd weight. PF$(n,n)$ contains exactly one sequence of length $2^n$ in $G_n$, i.e., a span $n$ de Bruijn sequence.
Hence, its weight is always even. For all the other parameters a perfect factor in which all the sequences have odd weight could
potentially exist.
Unfortunately, whether such perfect factors exist is an open problem. By Theorem~\ref{thm:FactorEven}, if $n=2^k-1$, then
all the sequences in a PF$(n,k)$ have odd weight. For the other parameters we can use the same technique used in the
proof of Theorem~\ref{thm:FactorEven}; the details are left for the reader. In other words, we can start with a span $k$ de Bruijn
sequence whose linear complexity is $2^{k-1} +\delta$, where $k \leq \delta \leq 2^{k-1} -k -1$ and $\delta \neq k+1$, and apply
the inverse operator $\bD^{-r}$, $r>0$, until we reach sequences with linear complexity $2^n$, where $n \geq k$. The sequences obtained
will be of odd weight by Lemma~\ref{lem:oddW_comp}. Other than these two techniques we do not know how to construct perfect factors
whose sequences all have odd weights.

It is important to note that in all the constructions mentioned so far, if a sequence $\cS$ is contained in the
perfect factor $\cF$, then $\bar{\cS}$ is also a sequence in $\cF$ (this includes self-dual sequences),
unless $\cF$ contains exactly one sequence. By Lemma~\ref{lem:comp_SD} all these sequences
of length $2^k$ are self-dual if and only if they are of linear complexity $2^{k-1}+1$.
However, perfect factors without these properties can also be constructed.

\begin{example}
When $k=4$ the length of the sequences is $2^4=16$ and there exists a PF$(n,4)$ with sequences of even weight
for each $n$, $4 \leq n \leq 14$. There exists a PF$(n,4)$ with sequences of odd weight for ${n \in \{ 5,6,8,12,15\}}$.
For $n \in \{ 7,9,10,11,13,14 \}$ we do not know whether such a perfect factor exists.

When $k=5$ the length of the sequences is $2^5=32$ and there exists a PF$(n,5)$ with sequences of even weight
for each $n$, $5 \leq n \leq 30$. There exists a PF$(n,5)$ with sequences of odd weight for $6 \leq n \leq 14$ and $n \in \{ 16,21,22,24,28,31\}$.
For $n \in \{ 15,17,18,19,20,23,25,26,27,29,30 \}$ we do not know whether such a perfect factor exists.

\hfill\quad $\blacksquare $
\end{example}

The following result is a consequence from the discussion in Section~\ref{sec:perfect_factor}, and implies that the requirement
for a perfect factor in Construction~\ref{con:2D_CCR} can be always satisfied.
\begin{lemma}
For each $n$ and $k$ such that $k < n < 2^k$ there exists a PF$(n,k)$ $\cF$ whose cycles (sequences) are not self-dual
and $\cS \in \cF$ if and only if $\bar{\cS} \in \cF$.
\end{lemma}
\begin{IEEEproof}
The methods used to generate perfect factors in Theorem~\ref{thm:FactorEven} also allows us to generate perfect factors
without self-dual sequences, and if $\cF$ is the PF$(n,k)$, then $\cS \in \cF$ if and only if $\bar{\cS} \in \cF$.
This is apart for the case when $n=2^{k-1}$ and the sequences of PF$(n,k)$ have linear
complexity $2^{k-1}+1$. In this case by Lemma~\ref{lem:comp_SD} the obtained sequences are self-dual.

For the case $n=2^{k-1}$ we consider a span $k-1$ de Bruijn $\cS$ with linear complexity $c(\cS)=2^{k-1}-1$ which exists by
Lemma~\ref{thm:deBcomp}. Using the inverse $\bD^{-1} \cS$ we obtain a factor in $G_k$ with two sequences of length $2^{k-1}$ and
linear complexity $2^{k-1}$. By applying $\bD^{-1}$ on these two sequences we obtain two self-dual sequences of length $2^k$ and linear
complexity $2^{k-1}+1$
in $G_{k+1}$. Applying $\bD^{-(2^{k-1} -k -1)}$ on these two sequences yields $2^{2^{k-1}-k}$ sequences of length $2^k$
in $G_{2^{k-1}}$, i.e., a PF$(2^{k-1},k)$ $\cF$ with sequences having linear complexity $2^{k-1}+1 + 2^{k-1} -k -1=2^k-k$ and therefore
$\cF$ has no self-dual sequences and $\cS \in \cF$ if and only if $\bar{\cS} \in \cF$.
\end{IEEEproof}

We continue by summarizing the parameters of the constructions which were presented.
For the direct constructions the parameters are presented in Table~\ref{tab:direct}.
For the recursive constructions the parameters are presented in Table~\ref{tab:recursive}.

\begin{table}[htbp]
\centering
\begin{tabular}{|c|c|c|c|c|c|c|}
\hline
construction      & parameters      & perfect factor     & constructed DBAC     & size          \\ \hline
Construction~\ref{con:2D_CSR}    & $m=2^t-1$  & PF$(n,k)$  & $(2^k,2^t;n,m)$ & $2^{nm-k-t}$  \\ \hline
Corollary~\ref{cor:join_SD}   & $1 \leq \ell$, $k <n<2^k$, $\ell +1 \leq t \leq n 2^\ell -k$ & PF$(n,k)$ no self-dual & $(2^k,2^t;n,2^\ell)$ & $2^{n2^\ell-k-t}$  \\ \hline
Construction~\ref{con:one_column}  & $1 \leq t \leq n-k$  & PF$(n,k)$  & $(2^k,2^t;n,1)$ & $2^{n-k-t}$  \\ \hline
Construction~\ref{con:FF_DB}  & $2 \leq n$, $m+1 \leq 2^t \leq 2^{nm}$  & PF$_{2^n}(m,t)$  & $(2^n,2^t;n,m+1)$ & $2^{n(m+1)-k-t}$  \\ \hline
\end{tabular}
\vspace{0.2cm}
\caption{Direct constructions of DBACs}
\label{tab:direct}
\end{table}

\begin{table}[htbp]
\centering
\begin{tabular}{|c|c|c|c|c|c|c|}
\hline
construction   & columns' weight      & parameters      & perfect factor     & constructed DBAC     & size          \\ \hline
Construction~\ref{con:PF_PMC_direct}       & even & $2\leq k,n,m$, $t=m-\ell$ & PF$(m,m-\ell)$  & $(2^k,2^{m-\ell};n+1,m)$ & $2^m \Delta$  \\
    & & $0 \leq \ell$, $m < 2^{m-\ell}$ & &  & \\ \hline
Construction~\ref{con:DB_even}       & even & $2 \leq k,n,t,m$ & PF$(m,m)$  & $(2^k,2^{m+t};n+1,m)$ & $\Delta$  \\ \hline
Construction~\ref{con:halfDB_even}       & even & $3 \leq m < 2^t$, $2 \leq k,n$ & PF$(m,m-1)$  & $(2^k,2^{m+t-1};n+1,m)$ & $2\Delta$  \\ \hline
Construction~\ref{con:DBodd_PMC_direct}   & odd & $3 \leq m < 2^t$, $2 \leq n < 2^k$ & PF$(m,t)$ no self-dual  & $(2^{k+1},2^m;n+1,m)$ & $2^{m-1}\Delta$  \\ \hline
Construction~\ref{con:lastODD}       & odd & $3 \leq m < 2^t$, $2 \leq n < 2^k$  & PF$(m,m-1)$  & $(2^{k+1},2^{m+t-1};n+1,m)$ & $\Delta$  \\ \hline
\end{tabular}
\vspace{0.2cm}
\caption{Recursive constructions with $(2^k,2^t;n,m)$-DBAC of size $\Delta$}
\label{tab:recursive}
\end{table}

What new parameters of DBACs can be obtained by the recursive construction? For example,
what DBACs with new parameters can be obtained using Construction~\ref{con:DB_PMC_direct}? Direct constructions can be used to form
$(2^k,2^t;n,m)$-DBACs, where the columns of the codewords are sequences from a PF$(n,k)$ say~$\cF$.
Theorem~\ref{thm:DB_PMC_direct} implies the existence of a $(2^k,2^t;n+1,m)$-DBAC. But, such a DBAC can be obtained
using a direct construction with a PF$(n+1,k)$ that can be obtained from $\bD^{-1} \cF$. So applying the recursion in this way does not yield DBACs
with new parameters. On the other hand we can transpose the DBACs obtained from the direct constructions
and use the recursion to obtained DBACs with new parameters. For this purpose, we have to examine the weight of the rows
in the direct constructions. To demonstrate the necessary analysis we concentrate on
Construction~\ref{con:2D_CCR}.

Construction~\ref{con:2D_CCR} is used to obtain $(2^k,2^{t+1};n,2^t)$-DBACs.
The codewords are $2^k \times 2^{t+1}$ matrices, where the $i$-th column
is the complement of the $(2^t+i)$-th column. In other words, each row is a self-dual sequence whose linear complexity is
$2^t+1$, which implies that its weight is even. Moreover, applying $\bD^{-\ell}$ on the sequences up to $\ell \leq 2^t -2$
yields sequences of even weight, so we will obtain DBACs with new parameters using
Construction~\ref{con:2D_CCR}. Applying $\bD^{-2^t+1}$ yields sequences with linear complexity $2^{t+1}$ which are
sequences of odd weight. This makes this construction very important for the recursive constructions.

Specifically, Let $\C$ be a $(2^k,2^{t+1};n,2^\ell)$-DBAC obtained by joining the codewords
of Construction~\ref{con:2D_CCR}. A codeword obtained in Construction~\ref{con:2D_CCR} is a $2^k \times 2^{t+1}$ doubly-periodic
array of the form $[X ~ \bar{X}]$, where $X$ is a $2^k \times 2^t$ array. Therefore, each row of the array is a self-dual sequence
whose linear complexity is $2^t +1$. Combining two such codewords yields an array of the form $[X ~ \bar{X} ~ Y ~ \bar{Y}]$.
Each row of such array is of length $2^{t+2}$ and its linear complexity is at most $2^{t+1} + 2^t$. If we continue to combine
codewords for a $(2^k,2^s;n,2^t)$-DBAC, where the linear complexity of a row in the obtained array is at most $2^s - 2^t$.
The codewords obtained can be transposed to obtain a $(2^s,2^k;2^t,n)$-DBAC whose columns have even weight
and linear complexity at most $2^s - 2^t$. On this code we can apply $j$ iterations of Construction~\ref{thm:DB_PMC_direct},
where $1 \leq j \leq 2^t -1$ to obtain a $(2^s,2^k;2^t+j,n)$-DBAC. The obtained codewords of the code can be transposed again to obtain
a new $(2^k,2^s;n,2^t +j)$-DBAC $\C'$. The codeword of $\C'$ can be merged by using the merge-or-split method to obtain
more DBACs with larger codewords and a smaller size.

Similar combinations of the direct and recursive constructions with a combination of the merge-or-split
method as described in Section~\ref{sec:join} yield more families of DBACs with various parameters.
A few families of $(2^k,2^t;n,m)$-DBACs obtained from the constructions are summarized in table~\ref{tab:family}.

\begin{table}[htbp]
\centering
\begin{tabular}{|c|c|c|c|c|c|c|}
\hline
DBAC     & parameters & constructions/reference          \\ \hline
$(2^k,2^t;n,1)$   & $k \leq n < 2^k$, $1 \leq t \leq n-k$  & Construction~\ref{con:one_column}    \\ \hline
$(2^k,2^t;n,2^\ell)$  & $1 \leq \ell$, $k \leq n < 2^k$, $\ell +1 \leq t \leq n 2^\ell -k$ & joining cycles/Corollary~\ref{cor:join_SD}  \\ \hline
$(2^k,2^s;n,2^t+j)$  & $1 \leq t$, $0 \leq j \leq 2^t-1$, $k \leq n < 2^k$, $t +1 \leq s \leq n 2^t -k$ & joining cycles/Corollary~\ref{cor:join_SD}/Construction~\ref{con:PF_PMC_direct}   \\ \hline
\end{tabular}
\vspace{0.2cm}
\caption{Infinite families of DBACs}
\label{tab:family}
\end{table}

\section{Conclusion and Future Work}
\label{sec:conclusion}

We have defined the concept of a de Bruijn array code, which generalizes the concept of a de Bruijn array.
Necessary conditions for the existence of such code were given in Lemma~\ref{lem:condPMC}.
The first goal in investigating these codes is to find whether the necessary conditions are also sufficient.
In other words, given four integers, $k$, $t$, $n$, $m$, such that $\Delta = 2^{nm-k-t}$, does there exist
a $(2^k,2^t;n,m)$-DBAC with $\Delta$ codewords? We conjecture that such a code exists, i.e., the necessary conditions are sufficient.

%If $\Delta =1$ the problem was completely solved by Paterson~\cite{Pat94}. The direct constructions together with the
%merge-or-split method to merge codewords and the recursive construction answer the existence question for some sets of parameters.
%While these constructions were
%designed for the case where a PF$(n,k)$ exists, i.e., when $k \leq n < 2^k$, i.e., $n < 2^k \leq 2^n$, by the transposing
%the codeword and applying the recursive constructions other sets of parameters can be obtained.

Several direct and recursive constructions were presented. These constructions yield de Bruijn array codes
with various parameters and several infinite families of such codes.
It should take more careful analysis on the constructed codes of this work to characterize exactly which classes of codes are
obtained and what is required to obtain in order to solve the conjecture.
To prove that the necessary conditions are indeed sufficient more constructions are required.
These are left for future research. Indeed, the range of possible parameters is so wide that
it will probably take some time before the conjecture is resolved.

Analogously to de Bruijn array codes, we can define pseudo-random array codes, a generalization of
pseudo-random arrays.
MacWilliams and Sloane~\cite{McSl76} defined pseudo-random arrays as follows.
\begin{definition}
A {\bf \emph{pseudo-random array}} $\cA$ is an $r \times s$ doubly-periodic array
such that each $n \times m$ nonzero matrix is contained exactly once
as a window in the array. Moreover, if $\cA$ is such an array, then
$\cA + \cA'$, where $\cA'$ is a nontrivial shift of $\cA$  and the addition is performed bit-by-bit, is another nontrivial shift of~$\cA$.
This is a shift-and-add-property.
\end{definition}
%Similarly, we have the following type of arrays.
%\begin{definition}
%A {\bf \emph{shortened perfect map}} (or a {\bf \emph{shortened de Bruijn array}}) is an $r \times s$ binary
%doubly-periodic array, where ${rs=2^{nm}-1}$, such that each nonzero $n \times m$ matrix appears exactly
%once as a window in the array.
%\end{definition}

The definition of pseudo-random arrays is generalized to pseudo-random array codes and will
be considered in the follow-up paper~\cite{CEL24}. They were also considered in a preliminary version
that discussed these array codes~\cite{Etzion24}. In the current paper the constructions of DBACs are based on combinatorial techniques
and the constructed codes are nonlinear.
In the follow-up paper~\cite{CEL24} the constructions of pseudo-random array codes will be based on algebraic techniques
and the constructed codes will be linear.

There are many directions to continue this research including questions on different types of codes. For example,
joining codewords from the codes obtained from Construction~\ref{con:2D_CSR}
leads to two open questions associated with an ordering of codewords of a given code $\C$.

Let $Q_4$ be the set defined by
$$
\cQ_4 \triangleq \{ (i_1,i_2,i_3,\ldots,i_{2^t}) ~:~ 0 \leq i_\ell \leq 2^{n-k} -1, ~ 1 \leq \ell \leq 2^t, ~ \sum_{\ell=1}^{2^t} i_\ell \equiv 1 ~ (\mmod ~ 2^{n-k}) \} .
$$
\begin{enumerate}
\item Order the elements of $Q_3$ in a cyclic list such that any two consecutive elements
in the list differ in exactly two adjacent coordinates.

\item Order the elements of $Q_4$ in a cyclic list such that any two consecutive elements
in the list differ in exactly two adjacent coordinates.
\end{enumerate}

\section*{Acknowledgement}

Part of this work was presented at the \emph{IEEE International Symposium on Information Theory}, Athens, Greece, July 2024.
This research was supported in part by the Israeli Science Foundation grant no. 222/19.
The author is indebted to two anonymous reviewers whose comments considerably improved the presentation of the paper.

%\appendix

%The appendix (or appendices) are optional. For reviewing purposes
%additional 5~pages (double-column) are allowed (resulting in a maximum
%grand total of 10~pages plus one page containing only
%references). These additional 5~pages must be removed in the final
%version of an accepted paper.

\end{document}